\documentclass[12pt,preprint]{aastex}
\usepackage{natbib}
\begin{document}

\title{Spatially Resolved Spectroscopy of Sub-AU-Sized Regions of T Tauri and 
Herbig Ae/Be Disks}

\author{J.A. Eisner\altaffilmark{1,2,3}, J.R. Graham\altaffilmark{2}, 
R.L. Akeson\altaffilmark{4}, 
%E.R. Ligon\altaffilmark{3},
%M. M. Colavita\altaffilmark{3}, 
\& J. Najita\altaffilmark{5}}

\altaffiltext{1}{Steward Observatory, University of Arizona, Tucson, AZ 85721}
\altaffiltext{2}{University of California at Berkeley, 
Department of Astronomy, 601 Campbell Hall, Berkeley, CA 94720}
\altaffiltext{3}{Miller Institute for Basic Research in Science, 
Berkeley, CA 94720}
\altaffiltext{4}{California Institute of Technology, 
Michelson Science Center MC 100-22,
Pasadena, CA 91125}
%\altaffiltext{3}{Jet Propulsion Laboratory, California Institute of Technology,
%Pasadena, CA 91109}
\altaffiltext{5}{National Optical Astronomy Observatory, 950 N. Cherry Avenue, 
Tucson, AZ 85719}


\begin{abstract}
We present spatially resolved near-IR spectroscopic observations of 15 young 
stars.  Using a grism spectrometer behind the Keck Interferometer, we
obtained an angular resolution of a few milli-arcseconds and a spectral 
resolution of 230, enabling probes of both gas and dust in the inner disks
surrounding the target stars.  
We find that the angular size of the near-IR emission typically  
increases with wavelength, indicating hot, presumably
gaseous material within the dust sublimation radius.  Our data also 
clearly indicate Br$\gamma$ emission arising from hot hydrogen gas,
and suggest the presence of water vapor and carbon monoxide gas in the 
inner disks of several objects.  This gaseous emission is more compact
than the dust continuum emission in all cases.  We construct simple physical
models of the inner disk and fit them to our data to constrain the spatial 
distribution and temperature of dust and gas emission components.
\end{abstract}

\section{Introduction \label{sec:intro}}
Protoplanetary disks provide a reservoir of material from which planets may 
form, and the abundance and properties of extra-solar 
planets \citep[e.g.,][]{MARCY+05}, 
as well as the architecture of our own solar system,
suggest that planets frequently form in or migrate through inner regions of 
protoplanetary disks.  Furthermore, the innermost disk regions represent the 
interface between the inwardly accreting disk and the magnetized central star,
and it is here that material accretes inward or is launched in winds or
outflows \citep{SHU+94}.  Knowledge of the distribution of material in the 
inner disk is therefore crucial for understanding  the 
mass assembly and angular momentum evolution of pre-main-sequence stars.

Modeling of spectral energy distributions \citep[e.g.,][]{BBB88,LA92}
and spectrally resolved gaseous emission lines 
\citep[e.g.,][]{NAJITA+96,BB04,NAJITA+06} provide important insights into the 
structure of protoplanetary disks in the terrestrial planet forming region.
However, since these techniques typically 
rely on spectral information as a proxy for
spatial information, they require assumptions about underlying geometric,
temperature, or velocity structure.  Recently, the technique of 
spectro-astrometry, which capitalizes on the fact that emission centroids
can be measured more precisely than the available angular resolution,
has enabled less ambiguous constraints on disk structure \citep{PONTOPPIDAN+08}.
However, substantial gaps in our understanding remain due to a lack of high 
angular resolution observations.

Near-IR interferometry, which synthesizes a large aperture using
two or more smaller, separated apertures, can achieve orders of magnitude
higher angular resolution than conventional telescopes, and can
spatially resolve disk terrestrial regions.  For example, the Keck 
Interferometer, which combines the light from the two 10-m Keck apertures
over an 85-m baseline, achieves a resolution approximately an order of 
magnitude higher than that attained with a single aperture.  
This means that angular scales of a few milli-arcseconds,
corresponding to a few tenths of an AU at typical distances to nearby
star-forming regions, can be spatially resolved with near-IR interferometers.

Most interferometric measurements to date have probed only inner disk dust 
\citep[e.g.,][and references therein]{MILLAN-GABET+07}, which typically 
dominates the near-IR emission.  To distinguish gas and dust, spectrally
dispersed observations are required. Very few sources have been 
observed with spectrally dispersed interferometric observations to date
\citep{EISNER+07a,EISNER07,MALBET+07,TATULLI+07,KPO08,ISELLA+08}.  
The observations showed 
intriguing evidence that gas and dust are not distributed
uniformly in inner disk regions.  Moreover, while observations of stars less
massive than a few M$_{\odot}$ found gaseous emission to be more compact than
dust emission \citep{EISNER+07a,EISNER07}, observations of the Br$\gamma$
emission line around two more luminous stars found the gas to be
more extended than the dust, presumably because the Br$\gamma$ emission traces 
outflows from
these young systems \citep{MALBET+07,TATULLI+07}.  A larger sample is
required to further investigate such potential trends, and to
constrain the general properties of inner disk gas in young stars.

Here we present spectrally dispersed near-IR interferometry observations of
a sample of young stars, including four T Tauri stars and 11 
Herbig Ae/Be stars.  Our data
constrain the relative spatial and temperature distributions of dust and
gas in sub-AU-sized regions of the disks around these stars.  
%In several cases,
%we determine the relative distributions of different gaseous species, 
%including hydrogen, CO, and H$_2$O.

\section{Observations and Data Reduction \label{sec:odr}}

\subsection{Sample \label{sec:sample}}
We selected a sample of young stars (Table \ref{tab:sample})
known to be surrounded by protoplanetary
disks, all of which have been observed previously at near-IR
wavelengths with long-baseline interferometers 
\citep{MST01,EISNER+04,EISNER+05,EISNER+07c,COLAVITA+03,MONNIER+05,
AKESON+05b,AKESON+05}.  All targets except one have been previously
spatially resolved in the near-IR.  HD 141569, which has not been
resolved, is thought to possess a disk with a cleared inner region; however
we include the source here to investigate whether any spectral features
are spatially resolved even though the continuum emission is not.

Our sample includes four T Tauri stars, pre-main-sequence analogs of solar-type
stars like our own sun, and 11 Herbig Ae/Be stars, 2--10 M$_{\odot}$ 
pre-main-sequence stars.  The main selection criterion in choosing this
sample is source brightness (and because Herbig Ae/Be stars are typically
brighter than T Tauri stars, our sample has more of the former).  
Our experimental setup
imposes limiting magnitudes of $K \sim 7$ at near-IR wavelengths and 
$V \sim 12$ at optical wavelengths.
We also require that sources be at zenith angles of less than 
$\sim 50^{\circ}$, which excludes from our sample
any sources with $\delta \la -35^{\circ}$. 
Our sample includes most of the sources in the Herbig-Bell catalog 
\citep{HB88}, as well as several additional young stars discovered elsewhere,
that meet our selection criteria.

\subsection{Observations \label{sec:obs}}
We obtained Keck Interferometer (KI) observations of our sample on 
UT 2006 November 12 and UT 2007 July 3.
KI is a fringe-tracking long baseline near-IR Michelson
interferometer combining light from the two 10-m Keck apertures 
\citep{CW03,COLAVITA+03}.  Each of the 10-m apertures is equipped with
an adaptive optics (AO) system that corrects phase errors caused by
atmospheric turbulence across each telescope pupil, and thereby maintains 
spatial coherence of the light from the source across each aperture.  The
AO systems require sources with $V$ magnitudes brighter than $\sim 12$.
Optical beam-trains transport the 
light from each aperture through an underground tunnel to delay lines,
beam combination optics, and the detector.

The two outputs of the beam combiner are sent into a dewar that contains a 
HAWAII array. Interferometric fringes are measured by modulating the relative 
delay of the two input beams and then measuring the modulated intensity level 
of the combined beams during four ``ABCD'' detector reads \citep{COLAVITA99}.
The measured intensities in these reads are also used to measure
atmosphere-induced fringe motions, and a servo loop removes these motions
to keep the fringe centered near zero phase.

KI normally measures science data in a spatially filtered wide-band channel,
with a second output of the beam combiner, dispersed over four
channels using a prism, used for group delay tracking.  
For the observations discussed here, we used a mode where the prism is 
replaced with a grism
providing an order of magnitude higher dispersion, and we used this
spectrally dispersed output for our science measurements.  
The grism, whose properties are described in \citet{EISNER+07b}, 
provides a spectral resolution of $R=230$, with 42 
10-nm-wide channels across the $K$-band. To obtain adequate signal-to-noise
for the group delay measurement with our grism, which has 10 times more pixels
and hence 10 times more read noise than the prism that is normally used, 
we require a star brighter than $K \sim 7$.

In the following sections we discuss the calibration, modeling, and
interpretation of our $V^2$ and flux data.  We exclude
HD 141569 and VV Ser from most of this discussion.  HD 141569 is unresolved 
across our observing band, and thus we can say only that the source is
small compared to the fringe spacing at all observed wavelengths.
Conversely, we were unable to measure fringes for VV Ser (despite measuring
strong $K$ band flux), and for this object we can say only that the
angular size is large compared to the fringe spacing. 
We defer discussion of these sources to \S \S 
\ref{sec:hd141569}--\ref{sec:vvser}.

\subsection{$V^2$ Calibration \label{sec:v2cal}}

We measured squared visibilities ($V^2$) for our targets and calibrator
stars in each of the 42 spectral channels provided by the grism. 
The calibrator stars are main
sequence stars, with known parallaxes,
whose $K$ magnitudes are within 0.5 mags of the target $K$ magnitudes
(Table \ref{tab:sample}).
The system visibility (i.e., the point source response of the interferometer)
was measured using observations of these calibrators, whose angular sizes
were estimated by fitting blackbodies 
%(with the temperature constrained by the
%measured spectral type) 
to literature photometry.  These size estimates are not crucial since the
calibrators are unresolved (i.e., their angular sizes are much smaller
than the interferometric fringe spacing) in almost all cases.  
HD 163955, a calibrator
for MWC 275, is mildly resolved; we account for this when computing the
system visibility.

We calculated the system visibility appropriate to each target scan
by weighting the calibrator data by the internal scatter and the temporal and 
angular proximity to the target data \citep{BODEN+98}.  For comparison, we 
also computed the straight average of the $V^2$ for all calibrators used for a 
given source, and the system visibility for the calibrator observations 
closest in time.  These methods all produce results consistent within the 
measurement uncertainties.  We adopt the first method in the analysis that 
follows.

Source and calibrator data were corrected for 
standard detection biases as described by \citet{COLAVITA99} and averaged into 
5-s blocks. We accounted for a known bias related to flux level
by applying an empirically determined 
correction\footnote{http://msc.caltech.edu/software/KISupport/dataMemos/fluxbias.pdf}.  
Calibrated $V^2$ were then computed by dividing the average
measured $V^2$ over 130-s scans (consisting of 5-s sub-blocks) for
targets by the average system visibility.  
Uncertainties are given by the quadrature 
addition of the internal scatter in the target data
and the uncertainty in the system visibility. 
We average together all of the calibrated data for a given source 
to produce a single measurement of 
$V^2$ in each spectral channel.  The observations of our targets typically
spanned $\la 1$ hour, and the averaging therefore has a 
negligible effect on the uv coverage.  
%Previous observations showed that
%this technique produces channel-to-channel uncertainties of a few percent
%\citep{EISNER+07b,EISNER07}.  

\begin{figure}
\plottwo{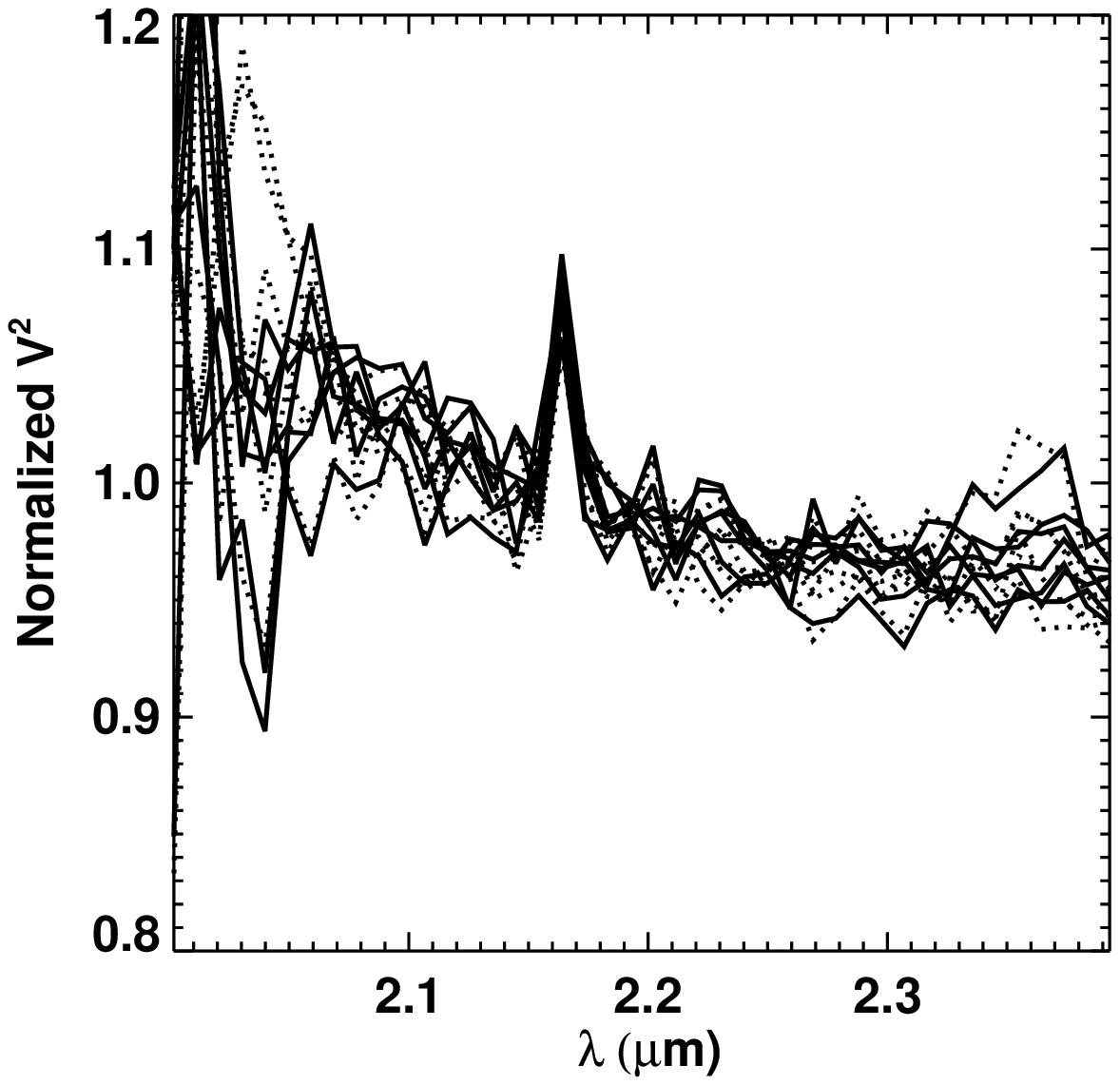}{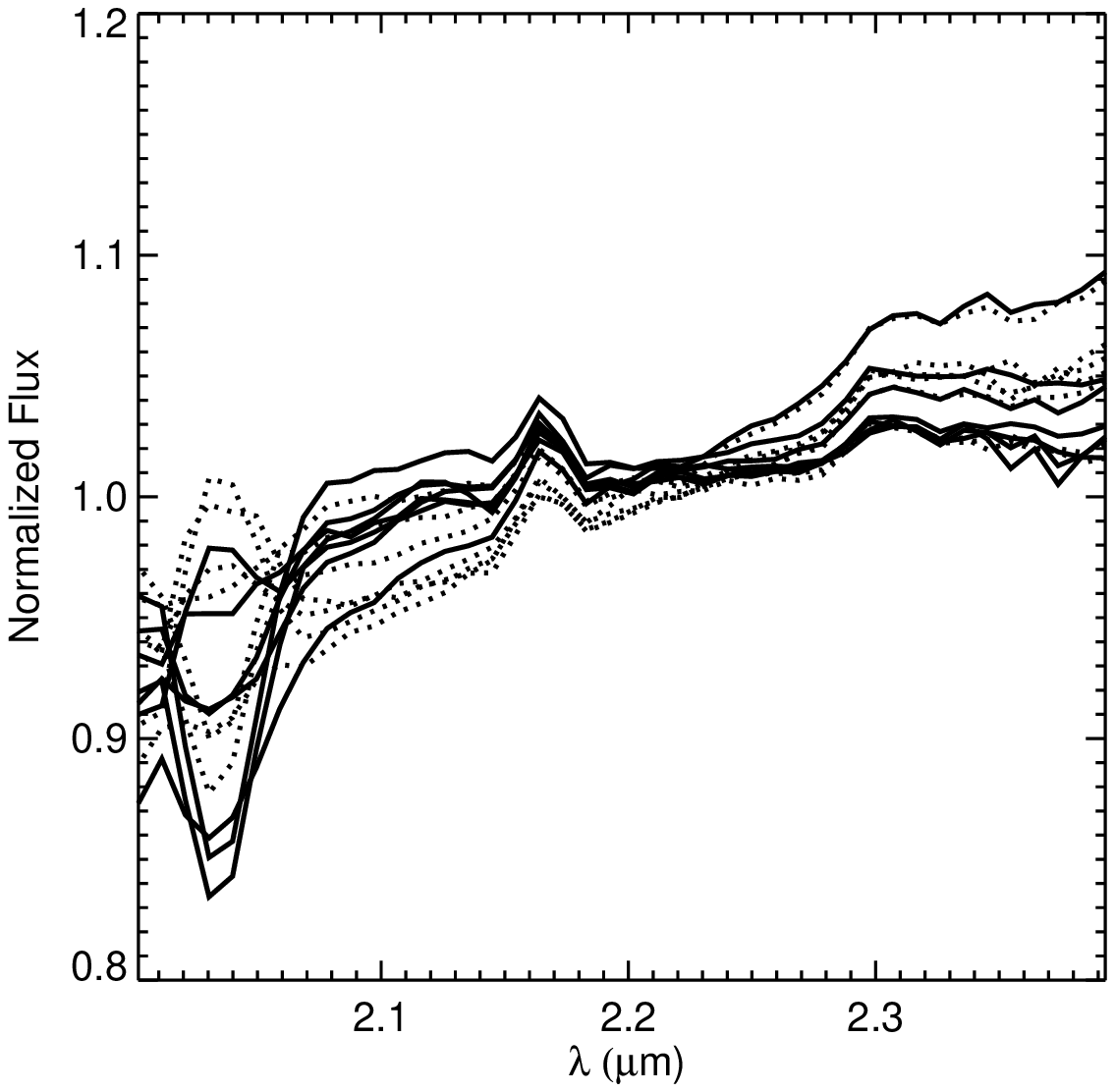}
\caption{$V^2$ ({\it left}) and fluxes ({\it right}) for several scans
on the target V1295 Aql.  Observations of a calibrator bracketing each target 
scan were used for calibration, and each line corresponds to one scan.
The solid curves show the results when HD 183324 was used as the
calibrator, and dotted lines show results when HD 188385 was used.
While we average all of these measurements together to determine the
average $V^2$ and fluxes for a target, we plot the results for each
scan and calibrator individually here to demonstrate statistical and
systematic uncertainties in the data.  Note that the strong feature
at 2.165 $\mu$m appears in all scans, and 
corresponds to Br$\gamma$ emission intrinsic to the source.
\label{fig:errors}}
\end{figure}

We investigate the uncertainties in our calibration procedure in
several ways.  First, we examine the calibrated $V^2$ for each scan for
one of our targets (V1295 Aql) where we obtained several scans.
We also compare calibrated $V^2$ using one calibrator or another.
Differences between scans and between the two calibrators provide an
estimate of the uncertainties. Results of this test are shown in Figure 
\ref{fig:errors}.  Across most of the band, channel-to-channel
uncertainties are a few percent or less.  However, there is a large dispersion
in the various measurements around 2.05 $\mu$m.  

A second probe of the uncertainties is provided by
applying our calibrations to two 
main-sequence check stars, HD 167564 and HD 171149 
(Figure \ref{fig:systematics}).
These stars were observed as calibrators for VV Ser, and are $30$--$40^{\circ}$
away from the other nearest calibrators in our dataset.  Thus, we 
expect the uncertainties derived here to be larger than for our target stars,
which are within 10$^{\circ}$ of their calibrators.
For these two stars, the standard deviation of the calibrated $V^2$ across
the bandpass is 1--2\%.  
However, the data for both stars exhibit an apparently systematic feature 
around 2.05 $\mu$m, the same spectral region that exhibits larger uncertainties
in Figure \ref{fig:errors}.  

\begin{figure}
\plotone{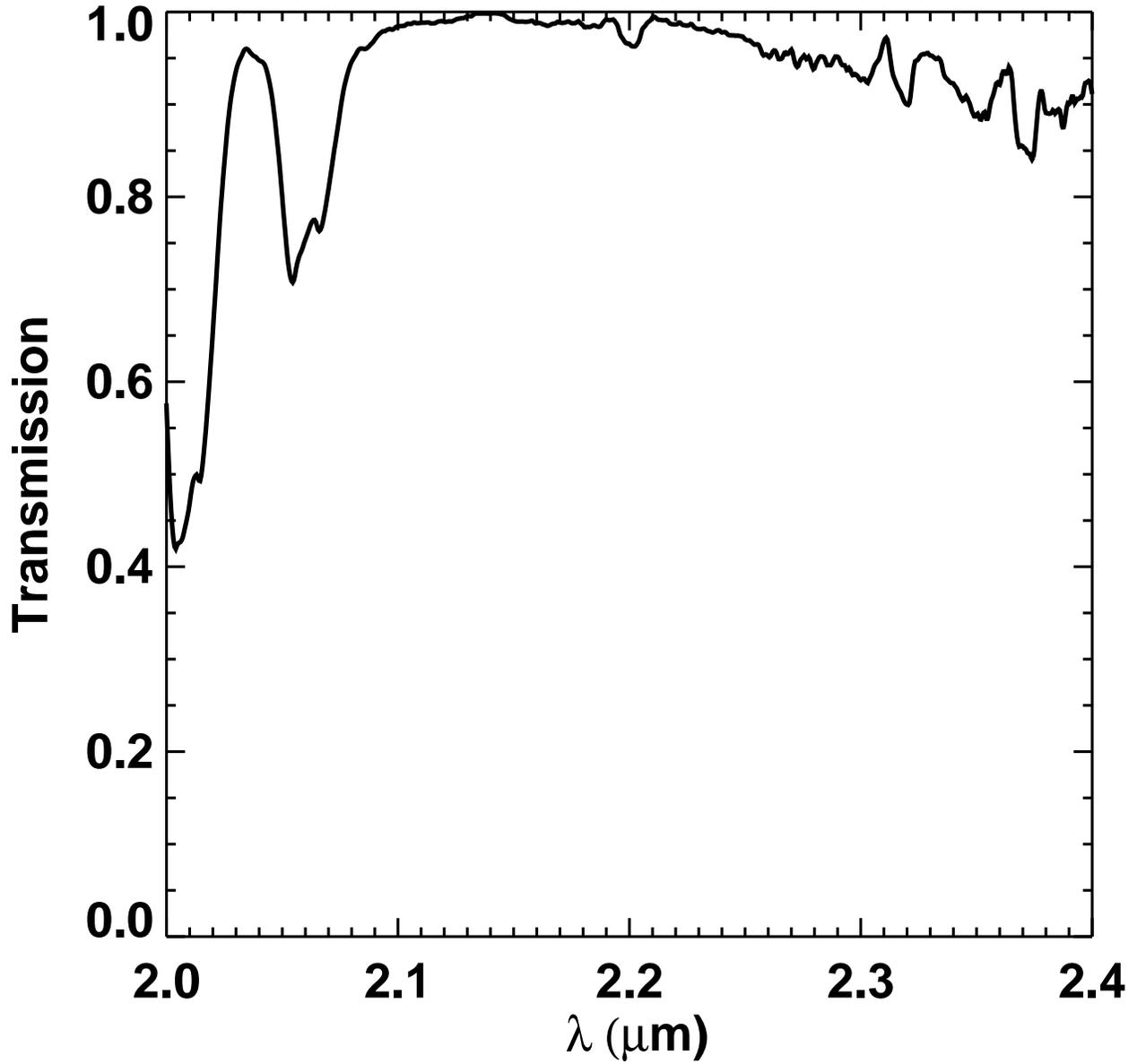}
\caption{The fractional transmission of the atmosphere over Mauna Kea
for 1.6 mm of precipitable water vapor and an airmass of 1.5.  The curve
has been smoothed to the spectral resolution of our observations.  The
strong absorptions on either side of 2.05 $\mu$m are due to atmospheric
CO$_2$.
\label{fig:transmission}}
\end{figure}

We do not have a simple
explanation for these large errors at the short-wavelength end of our band.
Telluric CO$_2$ features lead to absorption (and
hence lower photon counts) on either side of 2.05 $\mu$m, but not at 2.05
$\mu$m; a plot of the
atmospheric transmission is shown in Figure \ref{fig:transmission}.
Furthermore, it is difficult to imagine how any atmospheric
or instrumental absorption could vary as quickly as the observed variations
(and Figure \ref{fig:errors} shows that fluxes in this region 
vary both positively and negatively, contrary to expectations for 
time-variable absorption).
We speculate that the observed variability near 2.05 $\mu$m may be due
to constructive and destructive interference caused by a known 
phase irregularity at this wavelength in a dichroic optic. 

We adopt 3\% channel-to-channel uncertainties for our target stars.  However,
Figures \ref{fig:errors} and \ref{fig:systematics} indicate that data
shortward of $\sim 2.05$ $\mu$m likely has larger errors, and should be treated
with additional caution.  The normalization of $V^2$ versus wavelength 
(i.e., the average value of $V^2$ across the band) has an additional 
uncertainty.  Observations of binary stars with known orbits show that 
the calibrated $V^2$ have a systematic uncertainty of 
$\la 3\%$\footnote{http://msc.caltech.edu/software/KISupport/dataMemos/fluxbias.pdf}.  
We therefore assume that in addition to the 3\% channel-to-channel 
uncertainties described above, the normalization of $V^2$ is uncertain by 
$\sim 3$\%.

Calibrated $V^2$ for our sample are shown in Figure \ref{fig:data}.
As discussed above, HD 141569 and VV Ser are excluded from these plots
(and from our subsequent analysis) because they are unresolved and
over-resolved, respectively, in our observations.

\subsection{Flux Calibration \label{sec:fluxcal}}

We used the count rates in each channel observed during 
``foreground integrations'' \citep{COLAVITA99} to recover crude
spectra for our targets.  We divided the measured flux versus wavelength
for our targets by the observed fluxes from the calibrator stars, using
calibrator scans nearest in time to given target scans, and then multiplied
the results by template spectra suitable for the spectral types of
the calibrators.  

\begin{figure}
\plottwo{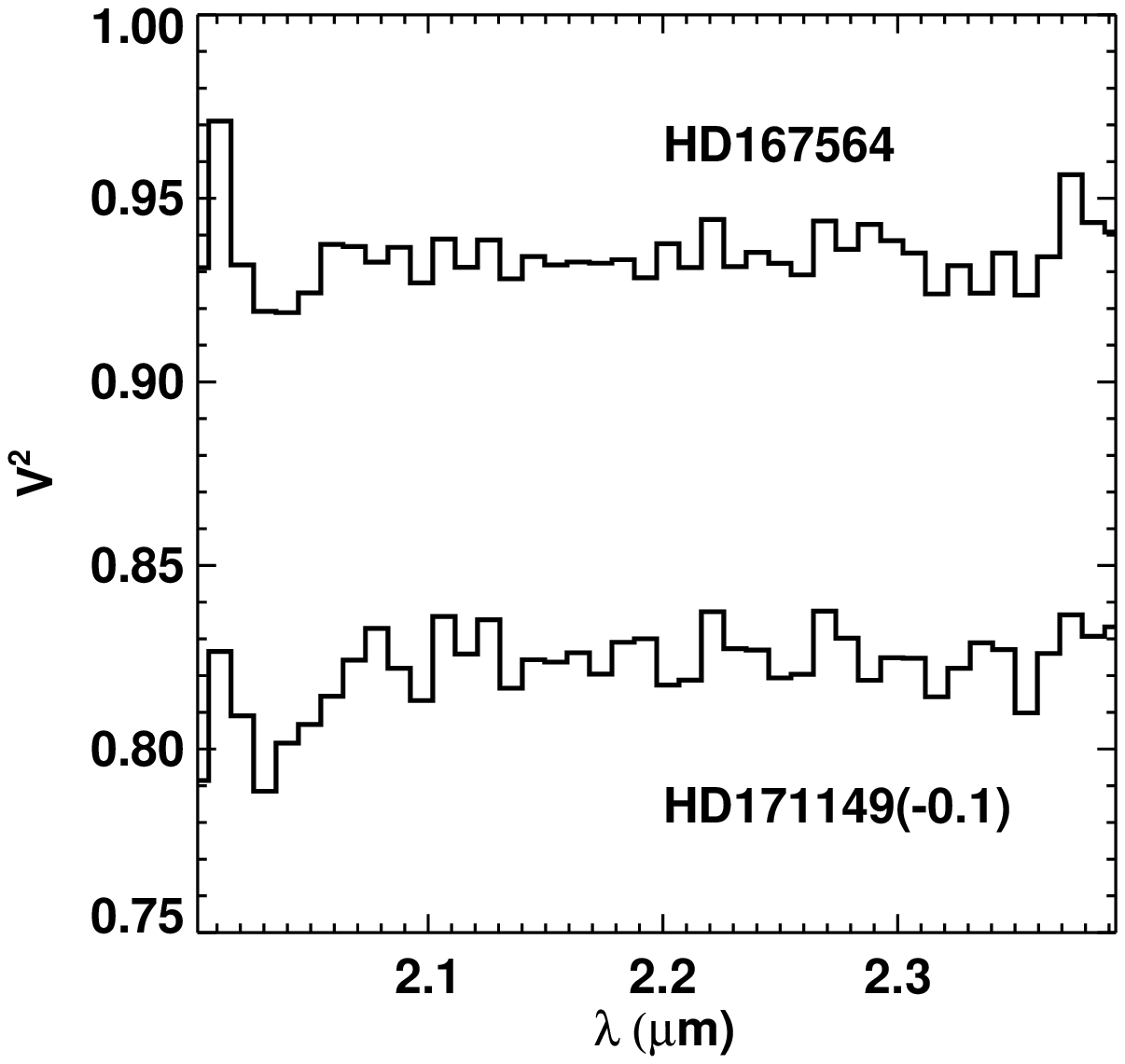}{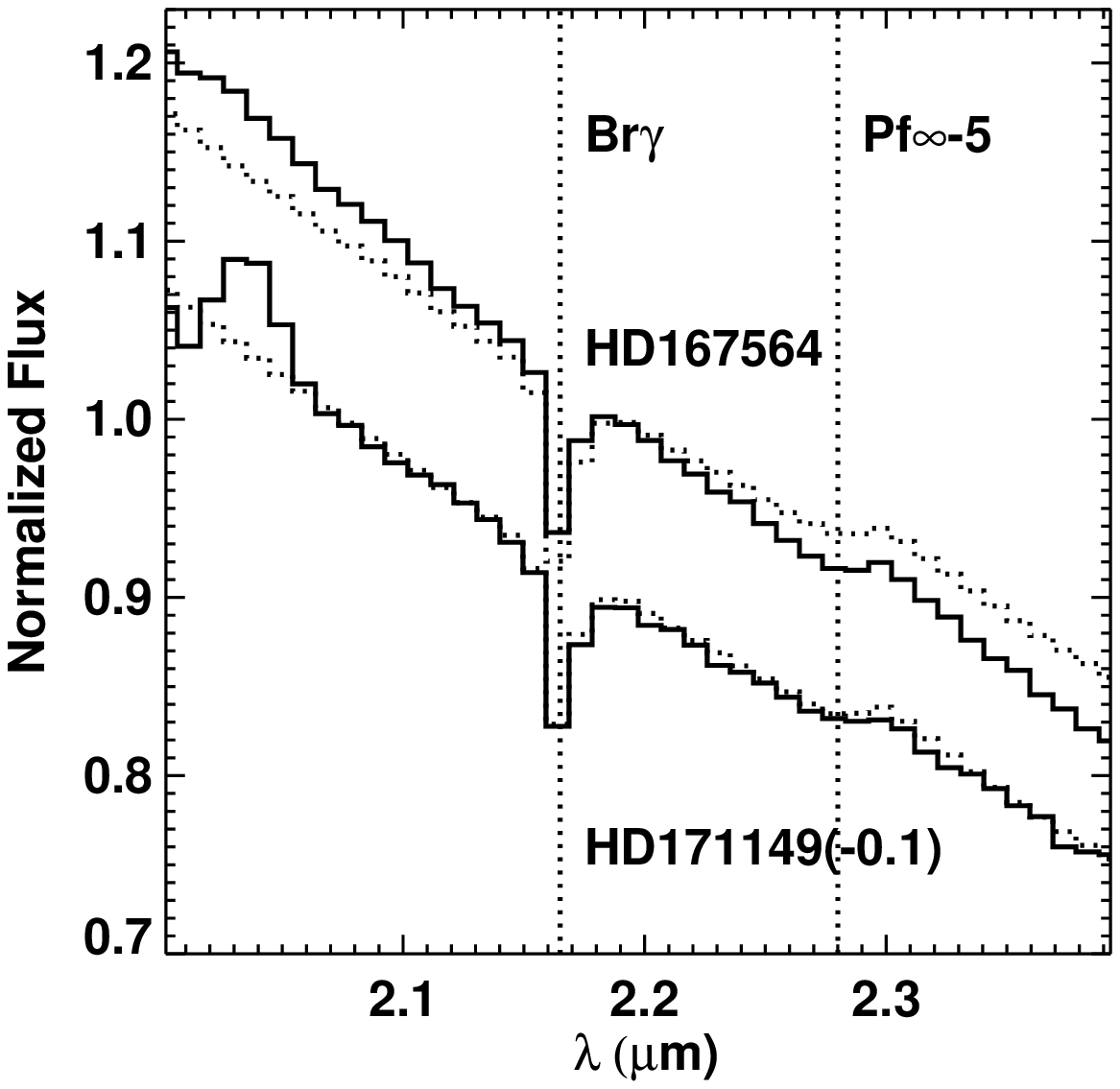}
\caption{Calibrated $V^2$ ({\it left}) and fluxes ({\it right}) for two
unresolved check stars.  The data for HD 171149 has been shifted down 
by 0.1 in both plots for ease of presentation.  We also plot the fluxes
predicted for these stars from Kurucz models ({\it dotted histograms}). 
We applied the same calibrations to these stars as we did to V1295 Aql.  
However, these stars are $\ga 30^{\circ}$ away from their calibrator stars,
substantially further than any of our targets are from their respective 
calibrator stars.  The calibration uncertainties seen here may therefore
be more severe than for any of our target data.
\label{fig:systematics}}
\end{figure}

We perform the same tests of our calibration procedure as employed
in \S \ref{sec:v2cal}.  Variations in calibrated fluxes for several
scans, using different calibrators, suggest that there are uncertainties
in the overall slopes of the spectra that lead to channel-to-channel 
uncertainties of $\sim 5$--10\% (Figure \ref{fig:errors}).
These slope variations may arise because of 
different coupling efficiencies of light at different wavelengths
into the single-mode fiber feeding the detector; the relative couplings
could also change with the atmospheric seeing.
Tests of this calibration procedure for
main sequence stars of known spectral type, calibrated using other
calibrator stars, indicate channel-to-channel uncertainties  of a few
percent (Figure \ref{fig:systematics}), and also show evidence for
these slope uncertainties.

Because our measured spectra have these potential slope errors, 
we use broadband photometry at near-IR wavelengths to infer the correct
normalization and slope.  We compiled near-IR
flux measurements from the literature 
\citep{MENDOZA66,MENDOZA68,GP74,HILLENBRAND+92,KH95,SKRUTSKIE+96,JM97,
MBW98,EIROA+02,KORESKO02,CUTRI+03,PGS03,EISNER+04}, 
including multiple measurements
where available, and then fitted a straight line to these data.  We
scaled our measured KI spectra so that the slope and normalization 
matched those of the fitted lines.  With these corrections, the spectra 
contain information about narrow spectral features, but do
not contain any original information about the overall normalization
or slope (this information comes from the broadband literature photometry). 
Moreover, since our targets tend to be photometrically variable, 
this procedure introduces some additional uncertainty in the absolute
flux level at the epoch of our KI observations.

We estimate uncertainties in our calibrated, slope-corrected
spectra of 5--10\%.  However, Figures \ref{fig:errors} and 
\ref{fig:systematics} both show evidence for larger errors around 
2.05 $\mu$m.  As for the $V^2$ calibration discussed in \S \ref{sec:v2cal},
the fluxes in this spectral region should be treated with caution.

\subsection{Features of Calibrated $V^2$ and Fluxes \label{sec:preresults}}
Figure \ref{fig:data} shows the $V^2$ and fluxes calibrated with the 
procedure outlined in \S \S \ref{sec:v2cal}--\ref{sec:fluxcal}.  The 
calibrated fluxes typically increase with wavelength across the
$K$ band, consistent with expectations for warm ($\la 2500$ K)
circumstellar emission.  The $V^2$ exhibit different behaviors with wavelength,
ranging from positive to negative slopes.  Interpretation of these
$V^2$ trends requires an accounting of the differing angular resolution
of the observations as a function of wavelength, since the resolution gets 
coarser at longer wavelengths.  It is more straightforward to explain
the $V^2$ behavior by first converting the measured $V^2$ into
angular sizes; we do this below in \S \ref{sec:uds}.

The fluxes and $V^2$ do not always behave monotonically across the observing
window, due to the presence of spectral features associated with
hot hydrogen gas, and warm CO and H$_2$O vapor.  
For example, MWC 480, MWC 275, and V1295 Aql
show clear bumps in both flux and $V^2$ near 2.165 $\mu$m, associated
with the Br$\gamma$ transition of hydrogen.  Spectral features like these may
arise in absorption in stellar photospheres, or in emission in the
circumstellar material.  In \S \ref{sec:mod}, we determine contributions
to the fluxes and $V^2$ from the central star, enabling investigation of the 
spectral features arising from the circumstellar emission.

\section{Modeling \label{sec:mod}}
In this section we use our $V^2$ and flux measurements to constrain the
distribution of dust and gas within 1 AU of our sample stars.  We first
consider the simplest possible model for the emission, a uniform disk
(\S \ref{sec:uds});
this model provides an estimate of how the angular size of the emission
depends on wavelength, with essentially no model assumptions.  We then
explore more physically realistic models that include both gas and
dust components.  Specifically, we model our sources as optically thin 
gaseous accretion disks that extend as
far as the dust sublimation radius, at which point emission from the
puffed-up dust sublimation front dominates.  We include various
gaseous species in the inner disk as needed to fit the data.
Because our $V^2$ and flux measurements contain contributions from both the 
circumstellar disks and the unresolved central stars of our targets, we remove
the stellar component from our measurements before fitting the models
described above.  The procedure by which stellar and circumstellar components
are separated is described in \S \ref{sec:ratios}.

\subsection{Separating Stellar and Circumstellar Components \label{sec:ratios}}
Before fitting physical models to the data, we remove the contribution
of the central stars from our measurements.  This enables us to model only the
circumstellar material around our sources.
Because the central stars are unresolved, we know that $V^2_{\ast}=1$, and
we need only the ratio of the stellar and circumstellar fluxes to
remove the stellar contribution to the measured $V^2$ and fluxes.

We estimate the circumstellar-to-stellar flux ratio at each observed
wavelength using spectral decomposition \citep[see, e.g.,][]{MST01}.  Using
optical photometry from the literature, we fit the stellar photosphere.
For each source we assume the spectral type given in the literature (also 
listed in Table \ref{tab:sample}) and determine
the stellar radius and reddening providing the best fit to the data; 
the reddening law of \citet{ST91} is used. 
In general we assume that the optical ($VRI$) photometry traces the
un-veiled stellar photosphere.  However, for AS 205 A, we have veiling
measurements at $R$ and $I$ bands from \citet{EISNER+05}, and we use
these to precisely fit the photospheric flux.  We then extrapolate the
photosphere to the $K$ band using Kurucz models, which include stellar
spectral features like Br$\gamma$ absorption in A stars or CO overtone
absorption in cooler stars.  We compare the extrapolated
stellar fluxes with our measured $K$-band fluxes to determine stellar and
circumstellar contributions.

With these circumstellar-to-stellar flux ratios, we remove the stellar
components of the visibilities:
\begin{equation}
V^2_{\rm disk} = \left\{\frac{\sqrt{V^2_{\rm meas}}(1+F_{\rm disk}/F_{\ast})-1}
{F_{\rm disk}/F_{\ast}}\right\}^2,
\label{eq:v2disk}
\end{equation}
where $F_{\rm disk}/F_{\ast}$ is the circumstellar-to-stellar flux ratio.
We fit our models for the circumstellar emission to 
$F_{\rm disk}$ and  $V^2_{\rm disk}$ below.

\subsection{Uniform Disk Sizes \label{sec:uds}}
\begin{figure}
\plotone{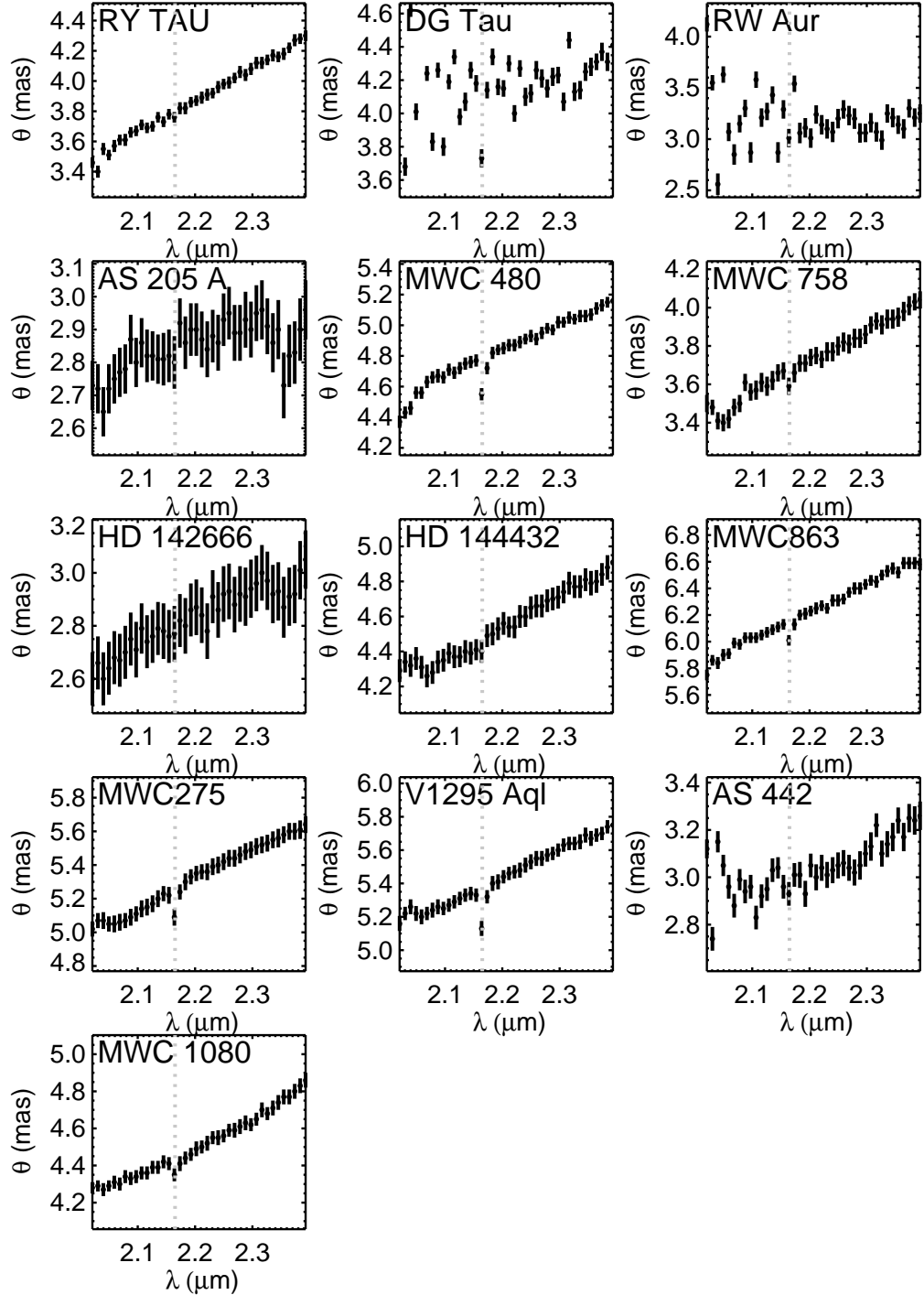}
\caption{Fitted uniform disk diameter as a function of wavelength for
our sample.  The dotted line in each panel marks the wavelength of the 
Br$\gamma$ transition of hydrogen.
\label{fig:udsizes}}
\end{figure}

Before interpreting our data in the context of physical models, we 
begin with a simple geometrical model: a uniform disk 
\citep[e.g.,][]{EISNER+03}.  We fit the
$V^2_{\rm disk}$ data for each source, in each channel, with the model.
The results, shown in Figure \ref{fig:udsizes}, 
give the ``spectral size distribution'' of the
$K$-band emission, illustrating how the spatial scale of the near-IR
emission depends on wavelength.

Figure \ref{fig:udsizes} shows that the angular diameter of the near-IR
emission appears to increase with wavelength for all sources in the sample
(although the trend is marginal for DG Tau and RW Aur, where the data 
have lower signal-to-noise).  As in previous work 
\citep{EISNER+07a,EISNER07,KPO08},
we interpret this slope as evidence for a hot emission component interior
to the inner edge of the dust disk.  This hot, compact component increasingly
dominates the observed emission at shorter wavelengths, and hence the
angular diameter of the emission appears to decrease.

In addition to showing increasing size with wavelength, Figure 
\ref{fig:udsizes} shows that several objects exhibit emission from
the Br$\gamma$ transition of hydrogen that is 
more compact than the circumstellar continuum emission.  
We see indications of compact Br$\gamma$ emission from MWC 480 
\citep[reported previously by][]{EISNER07},
MWC 758 (marginally), HD 144432 (marginally), MWC 863, MWC 275, 
V1295 Aql, and MWC 1080.

Finally, we see evidence of angular diameter changes due to CO opacity
in RW Aur.  Figure \ref{fig:rwaur} shows a zoomed-in view of the
spectral region encompassing several CO ro-vibrational overtone 
($\Delta v = 2$) bandheads.  The flux increases in each of the CO
bandheads, demonstrating the presence of CO emission in this object.
Moreover, the angular diameter appears to decrease within each of the
bandheads, indicating that the CO emission is more compact than the
continuum.

\subsection{Accretion Disk Models \label{sec:diskmods}}
As a more physically realistic model, we begin with a dusty circumstellar
disk whose emission is 
dominated by the hottest dust near the sublimation radius,
and then include 
gaseous emission interior to the inner edge of the dust.  We consider
several different gaseous opacity sources, including continuum emission,
Br$\gamma$ emission, water vapor, and CO.  For DG Tau and RW Aur,
we do not consider $V^2$ data measured shortward of 2.1 $\mu$m, since 
low fluxes ($<1$ Jy) lead to extremely noisy data at these wavelengths 
(Figure \ref{fig:data}).

We do not consider models that include only gaseous emission because
previous investigators have shown that ring-like dust sublimation fronts
are needed to fit broadband SEDs and interferometry data 
\citep[e.g.,][]{MUZEROLLE+03,EISNER+04}.
Furthermore, \citet{EISNER+07a} showed that gas-only disk models, while they 
can potentially reproduce observed spectrally dispersed visibilities,
can not simultaneously reproduce observed SEDs.  
We therefore do not consider gas-only models here.

We do not fit the broadband SED simultaneously with our 2.0--2.4
$\mu$m dataset here, to avoid additional complication.
Since our modeling is not constrained by SEDs outside of the $K$ band, the
inferred properties of the circumstellar emission may not perfectly reflect 
reality.  This is especially true for the gaseous emission, which is
often more compact than the spatial resolution of our observations.
Nevertheless, the modeling presented below demonstrates the presence
of gaseous emission components, and provides a rough estimate of the 
size scales and temperatures of these components.

\subsubsection{Dust Continuum Emission \label{sec:dust}}
We assume that the dust emission in our model arises from a single-temperature
ring of matter at the disk radius where temperatures become hot enough
for dust sublimation \citep[as in previous studies;
e.g.,][]{EISNER+04}.  This ring of emission approximates well \citep{EISNER+04}
the puffed-up inner edge expected for 
directly irradiated inner disks in hydrostatic equilibrium \citep{DDN01}.
The width of the ring is assumed to be two tenths of its radius, consistent 
with calculations of the expected width of the dust sublimation front 
\citep[e.g.,][]{IN05}.  The free parameters of this model are
the radius and temperature of the emission ring, $R_{\rm dust}$ and 
$T_{\rm dust}$.  We determine the best-fit values of these parameters
using a grid-based $\chi^2$ minimization.  Uncertainties are determined
directly from the $\chi^2$ surface \citep[e.g.,][]{EISNER+04}.

\subsubsection{Gaseous Continuum Emission \label{sec:gascont}}
We assume the radial temperature and surface density profiles of the gas in the
inner disk are described by power laws.  We take
$T(R) \propto R^{-1/2}$, as appropriate for optically-thin disk material
\citep[e.g.,][]{CG97}; $T(R)$ is the radial temperature profile and
$R$ is the stellocentric radius.  Details of the gaseous emission mechanism
may alter this relationship, but we will consider the radial temperature
law fixed for simplicity.
We assume that the surface density 
is described by $\Sigma(R) \propto R^{-3/2}$, as inferred for the protosolar
nebula \citep{WEID+77} and assumed in other disk models 
\citep[e.g.,][]{CG97,DDN01}.  The optical depth, $\tau$, is directly
related to the surface density, and hence $\tau_{\nu}(R) \propto R^{-3/2}$.
Depending on the normalization of the optical depth profile, the gas
may be entirely optically thin, entirely optically thick, or thick in
the innermost regions only.

We define the radial temperature and optical depth profiles over an inner disk
radius, $R_{\rm in}$, to an outer disk radius, $R_{\rm out}$.  
We take $R_{\rm out}$ to be the radius of the dust ring described in \S 
\ref{sec:dust}.  $R_{\rm in}$ is left as a free parameter.

The free parameters of the gaseous disk model are the inner radius,
$R_{\rm in}$, the temperature at the
inner radius, $T_{\rm in}$, and the optical depth at the inner radius,
$\tau_{\rm in}$.  When combined with the dust component, the
dust+gas continuum emission model has five free parameters.  We determine
the values of these parameters that provide the best-fit to the data using
a Marquardt-Levenberg algorithm.  Uncertainties on best-fit parameters
are computed from the covariance matrix.

%Figure \ref{fig:bblaw}) shows the expected flux versus stellocentric radius for
%this power-law disk model, where we have assumed a 9000 K stellar effective
%temperature to compute the normalization of the radial temperature profile.
If the gas is optically thick, then only the radial temperature profile
is important in determining the resultant flux.  However for most sources, 
simple disk models with optically thick gas predict fluxes larger by an order 
of magnitude than those observed from our sample.  We therefore expect
the gas to be (at least partially) optically thin in our modeling.
The 2 $\mu$m emission from optically thin gaseous disk models 
arises predominantly from the innermost radii. Models including
dust and optically thin gas thus 
resemble two-ring models \citep[like those considered in][]{EISNER+07a},
where hot gaseous emission appears in a ring near the inner radius and
warm dust emission arises from the ring-like dust sublimation front.

%If we assume we are in the Rayleigh-Jeans regime, and take 
%$T(r) \propto r^{-3/4}$, then 
%\begin{equation}
%F_{\rm annulus} \sim B_{\nu}(T(r)) \pi r^2 \sim T r^2
%\sim r^{-3/4} r^{2} \sim r^{5/4}.
%\end{equation}
%In contrast, for lower temperatures where we are not yet in the Rayleigh-Jeans
%limit, the temperature dependence of the blackbody law is much steeper, and
%the annular flux decreases with wavelength.  Even in the high-temperature
%regime, though, if the emission is optically thin, the flux may still
%decrease with radius.  The optical depth is proportional to the surface
%density: $\tau \propto \kappa \Sigma$.  Disk models suggest $\Sigma \propto
%r^{-1}$--$r^{-3/2}$, and thus if this term is included, the annular flux
%decreases with wavelength.  Figure \ref{fig:bblaw} 
%demonstrates what is happening.
%Note that the large fluxes arising from optically thick disks with such high
%temperatures is not compatible with the data for most objects, implying that 
%this gaseous emission must be (mostly) optically thin.  

\subsubsection{Water Vapor Emission \label{sec:h2omod}}
Several objects in our sample show additional flux, and accompanying
broad features in $V^2$, at wavelengths $\la 2.1$ $\mu$m.  These features
are (in some cases) broader than the systematic features discussed in 
\S \ref{sec:obs}, although the data is still
affected by these short-wavelength systematic effects.  In a previous
paper, we reported such features for MWC 480, and modeled the data by adding
a single-temperature ring with the opacity of water vapor to a two-ring
dust+gas model \citep{EISNER07}.  Here we refine this
simple model, and test whether the data for our sources
can be fitted with a model where water vapor emission occurs over a range of
radii from an inner radius where water is thermally dissociated out to the
dust sublimation radius.

We use opacities measured for water vapor  \citep{LUDWIG71}
at a range of temperatures
($\sim 1000$--3000 K) appropriate to protoplanetary disks.
Using the gaseous disk model described in \S \ref{sec:gascont}, 
we determine the optical
depth at each radius using the assumed surface density profile, which scales as
$\Sigma(R) \propto R^{-3/2}$: $\tau_{\nu}(R)=\Sigma(R) \kappa_{\nu}$.
For disk radii where $T<3000$ K, water vapor can exist and we use 
$\kappa_{\nu}$ from \citet{LUDWIG71}.  At smaller stellocentric radii, where
water would be dissociated, we retain the assumption of continuum emission.
While it is also possible that water emission, where is exists, adds to rather
than replaces the underlying continuum, we do not consider this possibility in 
our simple model.

%This continuum emission may arise from H$^-$ free-free interactions in the
%hot, dense inner disk (see \S \ref{sec:opacity} for further discussion).

\subsubsection{Brackett Gamma Emission \label{sec:brgmod}}
Figures \ref{fig:data} and \ref{fig:udsizes} show that Br$\gamma$ emission
signatures are present in the flux and $V^2$ data for several objects.
In fact, flat spectra in the vicinity of the Br$\gamma$ wavelength
(2.165 $\mu$m) may indicate the presence of Br$\gamma$ emission for
some objects.  Since more massive stars, including essentially all of the 
Herbig Ae/Be stars in our sample, have photospheric Br$\gamma$ absorption,
non-detection of this absorption feature in the spectra implies that
circumstellar Br$\gamma$ emission (and continuum excess) is filling
in the feature.  The effects of photospheric Br$\gamma$ absorption are included
in the Kurucz models we used to separate stellar and circumstellar components 
of the measured $V^2$ and fluxes (\S \ref{sec:ratios}).

We model circumstellar Br$\gamma$ emission by including additional
flux in the gaseous component of our models at 2.165 $\mu$m.  Since Br$\gamma$
traces very hot gas ($\sim 10^4$ K), we assume that the emission arises from 
the innermost disk annulus.  In fact, Br$\gamma$ emission may be even more
compact, but we can not constrain such small size scales with the available
angular resolution.  Inclusion of Br$\gamma$ emission in the model
introduces as an additional free parameter the gaseous line-to-continuum ratio 
in the innermost gaseous annulus.

\section{Results \label{sec:results}}

\subsection{Dust and Gas within 1 AU of young stars \label{sec:resmain}}
Our modeling shows clearly that single-temperature rings of emission
can not fit the data for our sample 
(Figure \ref{fig:fits}; Table \ref{tab:results}).  
This confirms previous inferences from spectrally dispersed interferometric
results \citep{EISNER+07a,EISNER07,KPO08,ISELLA+08} 
and from high angular resolution
continuum observations \citep{TANNIRKULAM+08}.
The implication is that models where all of the near-IR emission arises
from the inner edge of the dusty disk are untenable.  Rather, a warm
dusty component and a hotter, presumably gaseous, component at smaller
stellocentric radii are required to fit the data.

The simple dust+gas continuum model (\S \ref{sec:diskmods}) predicts $V^2$ and
fluxes compatible with the observations for most sources
(Figure \ref{fig:fits}).   
We also consider a model where the gaseous
opacity is due to water vapor at stellocentric radii where H$_2$O can
potentially exist.  As shown in Figure \ref{fig:fits}, these models
generally produce fits of comparable quality to the simple
continuum models.  Models including water vapor also produce more
physically realistic fitted values of $R_{\rm dust}$:. the fitted dust 
temperatures are between $\sim 1200$ and 1500 K, in agreement with expected 
sublimation temperatures for silicate dust in
protoplanetary disks \citep[e.g,][]{POLLACK+94}.  However, for RY Tau,
the simple dust+gas (continuum) model produces a superior fit to the data.  

\begin{figure}
\plotone{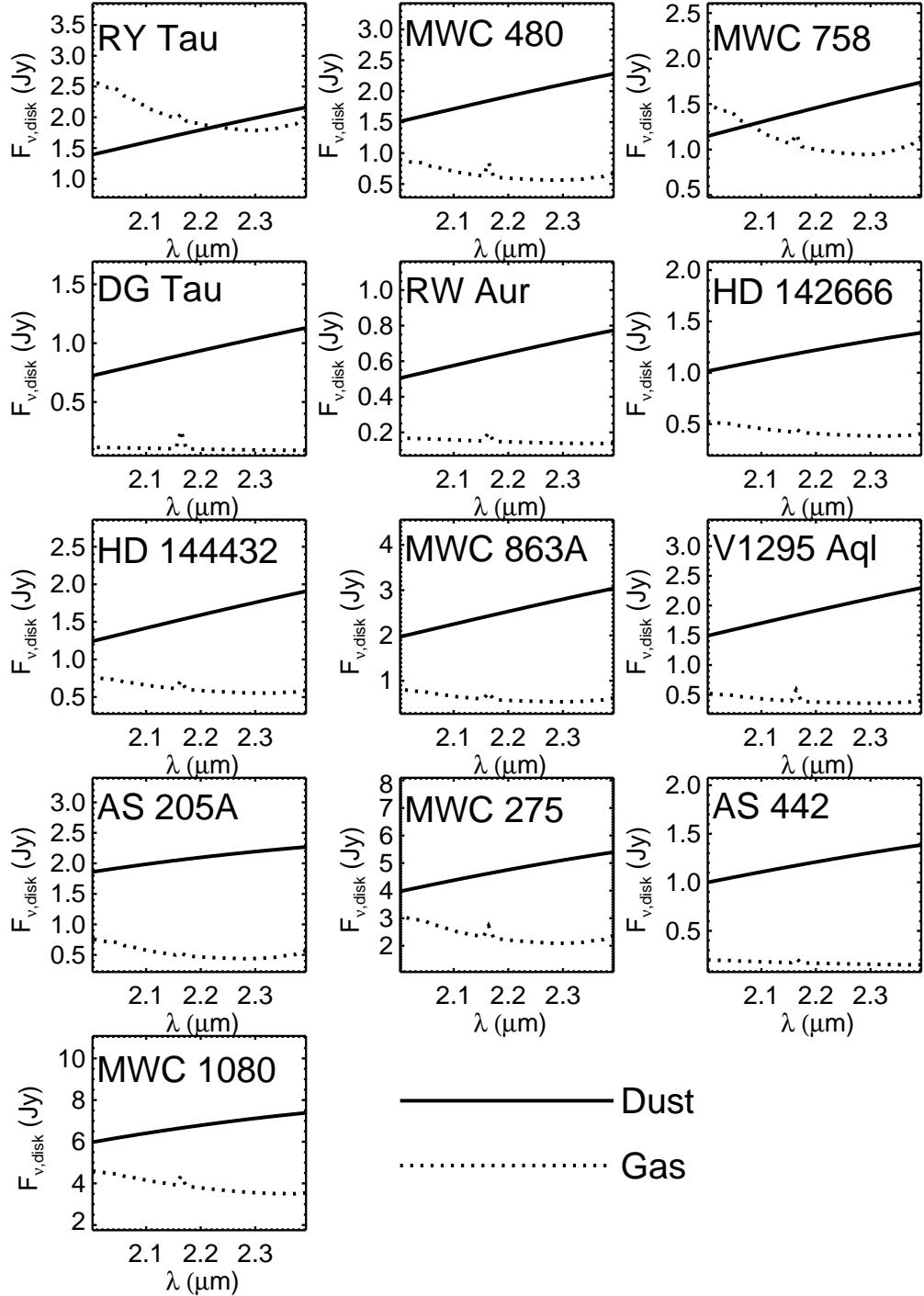}
\caption{Spectra of best-fit dust+water models, decomposed into dust 
({\it solid curves}) and gas components ({\it dotted curves}).  The parameters
of the models are given in Table \ref{tab:results}.  The gaseous component
includes continuum emission from material hotter than 3000 K, 
water vapor emission from gas cooler than 3000 K, and Br$\gamma$
emission.
\label{fig:components}}
\end{figure}

We illustrate the relative contributions of gas and dust to our models
in Figure \ref{fig:components}.  The figure shows the fluxes produced
by the gaseous component and the dust component for our best-fit models
including water vapor opacity.  The gaseous component includes continuum
emission, water vapor emission, and Br$\gamma$ emission.  In all cases, 
the gaseous flux represents a significant fraction of the total.

For all sources except AS 205 A, the best-fit
dust+water model requires emission from material substantially hotter
than 3000 K (Table \ref{tab:results}), implying the presence of substantial 
gaseous continuum emission.  The inferred inner disk temperatures for models
including water vapor opacity are, in most cases, higher than for models
assuming only gas continuum opacity.  
This difference suggests that the best-fit
for models including water vapor requires the gaseous emission to be dominated
by the hot continuum component rather than the cooler water vapor component.
For AS 205 A, in contrast, it appears that water vapor alone can explain the 
observed gaseous emission from the inner disk.

\begin{figure}
\plotone{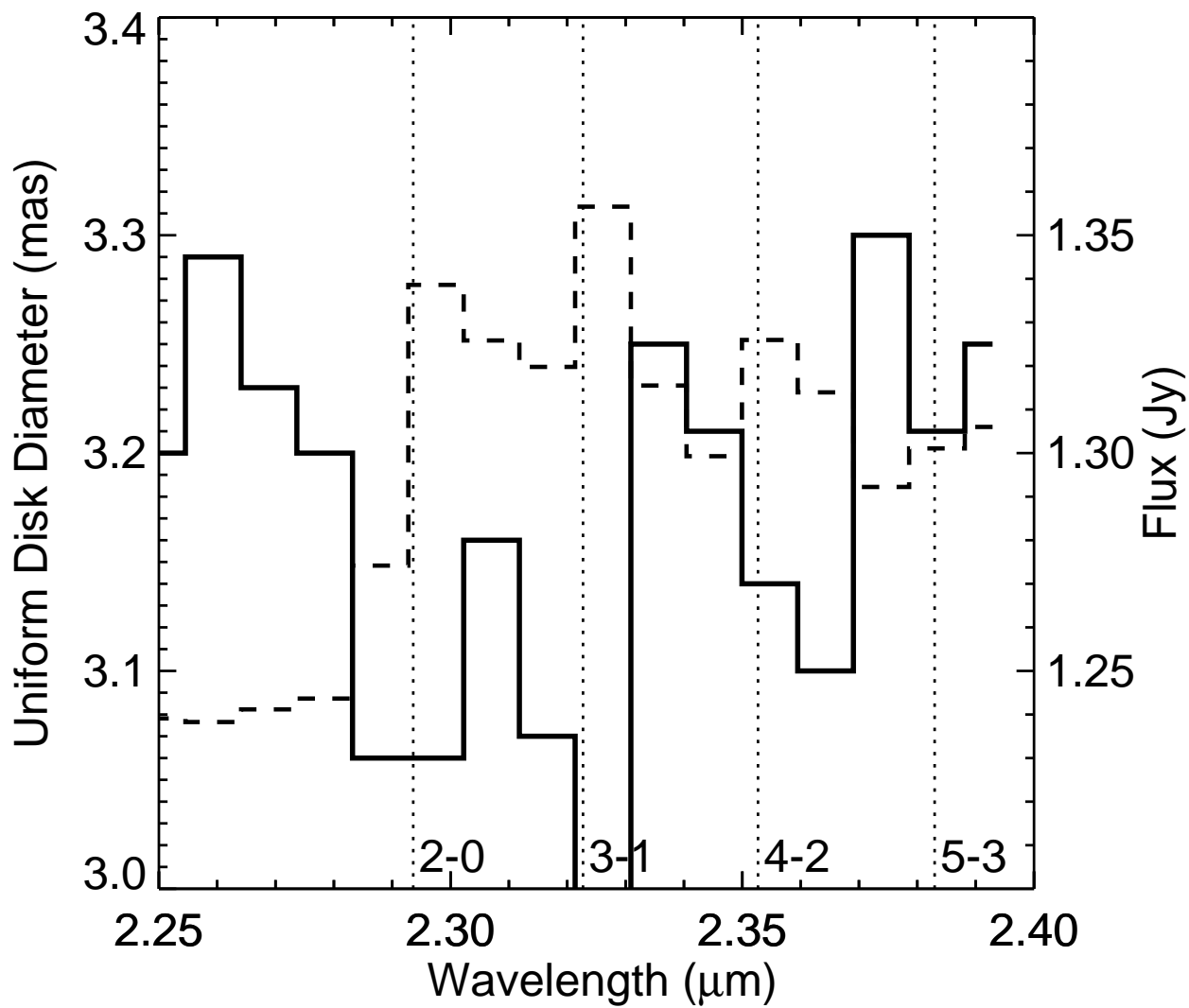}
\caption{Uniform disk diameter as a function of wavelength for RW Aur A 
({\it solid histogram})
in the spectral region of the CO overtone bandheads.  The measured fluxes
in this region are also plotted ({\it dashed histogram}).  
Dotted lines show the wavelengths of the CO overtone bandheads.  The uncertainties on measured sizes and fluxes, which can be found in Figures 
\ref{fig:udsizes} and \ref{fig:fits}, are $\sim 5\%$.
\label{fig:rwaur}}
\end{figure}

RW Aur shows evidence of spatially resolved CO in the inner
disk.  Emission from the $\Delta v=2$--0, 3--1, and 4--2 bandheads is
detected in our spectra, and appears to be more compact than the surrounding
continuum emission (Figure \ref{fig:rwaur}).  
This suggests an origin of the CO interior to the
dust sublimation radius.  The inferred size of the CO emission for this object
is compatible with that inferred from modeling of a high dispersion
spectrum in terms of a Keplerian disk model \citep{NCM03}.  

DG Tau,
which is also known to possess CO overtone emission \citep[e.g.,][]{CARR89},
shows no evidence of such emission in our data.  This discrepancy may be
due to the poor quality of our data for this object or to variability.
The CO emission from DG Tau is known to be variable from studies at multiple 
epochs \citep[e.g.,][]{CARR89,NAJITA+00,NCM03}, 
and some previous studies failed to detect the emission \citep[e.g.,][]{GL96}.

\begin{figure}
\plotone{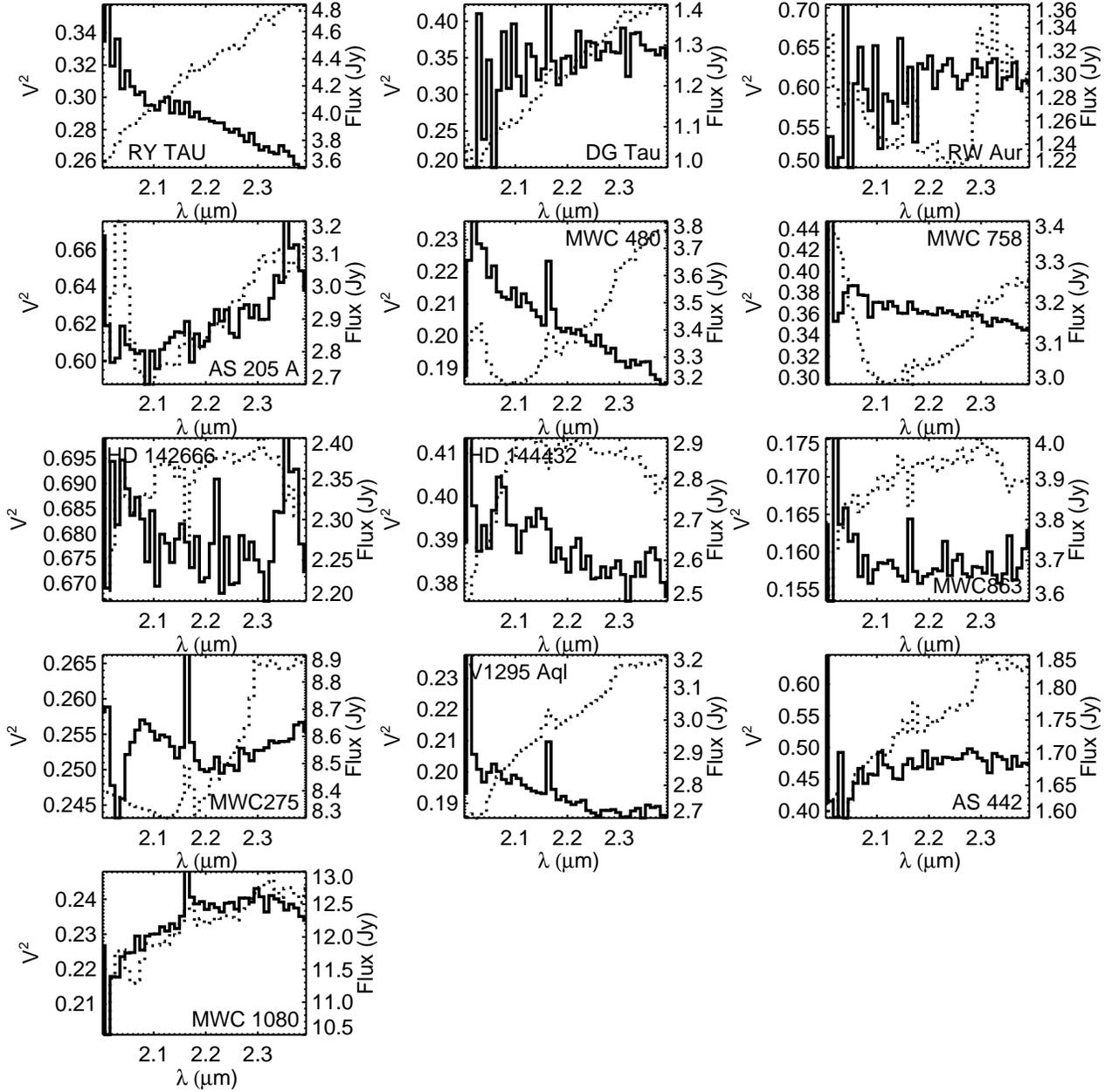}
\caption{Calibrated $V^2$ (solid histograms) and fluxes (dotted histograms)
for our sample.  Channel-to-channel uncertainties are estimated to be
3\% for $V^2$ and 5--10\% for fluxes.  For DG Tau and RW Aur, the noise in
the measured $V^2$ is larger for $\lambda \la 2.2$ $\mu$m, owing to the
low fluxes ($\la 1$ Jy) at these wavelengths.
\label{fig:data}}
\end{figure}

MWC 758, MWC 275, V1295 Aql, and AS 442 all show excess flux long-ward of
$\sim 2.3$ $\mu$m, consistent with the presence of CO overtone emission
(Figure \ref{fig:data}).  However, none of these show evidence in the
visibilities of size differences between CO emission and continuum.
Given the signal-to-noise and spectral resolution of our current data, we
do not attempt to model the CO emission from any of these targets here.
%We plan to re-observe these objects in the future with higher spectral
%resolution, which will provide higher signal-to-noise and enable tighter
%constraints on the relative distributions of CO and other gas and dust 
%in the inner disk.

Several objects exhibit Br$\gamma$ emission from hot hydrogen gas, and
this emission is more compact than the surrounding continuum emission in
all cases.  Table \ref{tab:brg} lists the inferred stellocentric radii from 
which the Br$\gamma$ emission originates.  However, we have somewhat 
arbitrarily placed the Br$\gamma$ emission at the innermost edge of
the gaseous accretion disk in our models (\S \ref{sec:brgmod}).  
Since the angular resolution of
our observations is $\sim 0.1$ AU, and there is some degeneracy in our models
between the Br$\gamma$ flux and angular scale, we can only state 
confidently that the Br$\gamma$ emission arises from radii less than 0.1 AU.

While our model fits indicate
that Br$\gamma$ emission is present around RW Aur, and is more compact 
than the continuum, we are not confident in this result given the large
uncertainties on $V^2$ for this source in this spectral region.
The data for DG Tau also show evidence for compact Br$\gamma$ emission,
but the noisy data in this spectral region argue for a cautious interpretation.
Both DG Tau and RW Aur have previously detected (spatially unresolved)
Br$\gamma$ emission \citep{FE01,NCT96}, providing some support for 
our tentative detections.  

MWC 480, HD 144432, MWC 863, V1295 Aql, and MWC 275,
all targets that exhibit compact 
Br$\gamma$ emission, were previously reported to show strong Br$\gamma$ 
emission in high-dispersion (spatially unresolved) spectroscopic observations
\citep{GARCIALOPEZ+07,BRITTAIN+07,BERTHOUD08}.  RY Tau, MWC 758,
HD 142666, and AS 205 also have previously reported Br$\gamma$ emission 
\citep{NCT96,FE01,GARCIALOPEZ+07,BERTHOUD08}.  However, the equivalent widths 
of the Br$\gamma$ emission are small relative to other sources 
(e.g., MWC 275 or DG Tau), and it is not surprising that we do not detect 
(or only marginally detect) Br$\gamma$ emission 
from these objects here.

\subsection{An unresolved source: HD 141569 \label{sec:hd141569}}
Previous near-IR interferometric observations of HD 141569 found it
to be unresolved in the $K$ band, suggesting that the $K$ band emission
is produced entirely by the unresolved central star.
We re-observed it here to search for spatially extended circumstellar emission 
associated with gaseous spectral features,
which might have been washed out in the previous, broadband observations.
The observations presented here, however, show that the source is
unresolved at all wavelengths.  We also see no evidence of flux above the
level expected from the stellar photosphere at any observed wavelength.
We conclude that the emission in each of the spectral channels included in our 
observations is more compact than $\sim 0.1$ AU (at the target distance),
as expected for stellar emission.  We may also infer that no spectral channels
contain emission brighter than $\sim 1\%$ of the stellar flux within the
50 mas field of view of KI (corresponding to stellocentric radii of 
$\sim 2.5$ AU for this target), since the incoherent contribution of any
stronger extended emission would have reduced the measured visibilities.
Our finding is consistent with previous
studies that found gaseous emission only at stellocentric radii $\ga 10$ AU
\citep{GOTO+06,BRITTAIN+07}, with no warm gas that could have been detected
in our observations.

\subsection{An over-resolved source: VV Ser \label{sec:vvser}}
One of the objects in our sample, VV Ser, appears to be over-resolved
in our observations.  That is, while we detect flux from the object, its
angular size is large compared to the $\sim 5$ mas fringe spacing and
hence its $V^2$ is un-detectably close to zero.  Because we can only place
a lower limit on the size of the emission, and can not trace how the
size depends on wavelength, we have excluded VV Ser from the analysis
presented above.

Previous observations of VV Ser, with fringe spacings comparable
to those obtained in the present study, 
measured a uniform disk angular diameter
of $\sim 4$ mas, and showed the disk to be nearly edge-on 
\citep{EISNER+03,EISNER+04}.  Our present non-detection indicates that 
the size scale of the $K$ band 
emission has varied significantly between 2003 and 2007.  
The source is also known the photometrically variable at optical
through infrared wavelengths \citep[e.g.,][]{HS99,EIROA+02}.

These findings may indicate variability associated with the UX Ori phenomenon.
In this scenario, a vertically extended inner edge of the nearly edge-on
disk around VV Ser periodically blocks the central star 
\citep[e.g.,][]{DULLEMOND+03,PONTOPPIDAN+07}.  When the star
is blocked, the near-IR emission would be dominated by the extended
circumstellar component, whereas when the star is visible, the observed
size would be the flux-weighted average of the unresolved star and the disk.

\section{Discussion \label{sec:disc}}

\subsection{Compact Br$\gamma$ Emission \label{sec:brgdisc}}
We observe Br$\gamma$ emission from several of our targets, and in all
cases this line emission appears more compact than the surrounding 
continuum.  This contrasts with previous results that found Br$\gamma$
emission more extended than \citep{MALBET+07} or on a comparable
spatial scale to \citep{TATULLI+07} the dust continuum.  Sources where
Br$\gamma$ appears more extended are high-mass stars, and \citet{EISNER07}
speculated that the Br$\gamma$ emission from young stars may trace both
infalling and outflowing components, with the latter increasingly dominating
for higher mass stars.

The results presented here belie this hypothesis.  We observe Br$\gamma$
emission more compact than dust continuum in sources spanning a range of
spectral types, from K3 to B0.  Based on our current findings,
we suggest that young stars typically produce most of their Br$\gamma$ emission
close to their central stars in accretion columns and/or shocks.  
Br$\gamma$ emission in extended winds seems rarely to be strong enough
to dominate the line emission.  However, given the small number of
Herbig Be stars observed to date, the picture is less clear for higher-mass 
($\ga 10$ M$_{\odot}$) stars.

\subsection{Inner Disk Gas: Trends with Stellar Luminosity \label{sec:gas}}
For the majority of our sources, models that include only dust emission
do not fit the data well, while models including hot gaseous emission
interior to the dust sublimation do fit the data (\S \ref{sec:resmain}).
However, if we examine only the less luminous stars in our sample, 
the T Tauri stars, we find that this conclusion is less robust.
For all of the T Tauri stars in our sample (RY Tau, DG Tau, RW Aur, 
and AS 205), dust-only models can provide relatively good matches to
the slopes seen for fluxes and $V^2$.  Models including gas and dust
are still superior, but the difference is not nearly as pronounced
as for the more massive Herbig Ae/Be stars in our sample.

We suggest a simple explanation for this trend.  For T Tauri sources,
the less luminous central stars provide less heating and hence the
dust can exist closer to the star than for the more massive Herbig stars.
While gaseous disks are likely to extend in to similar radii
in both types of sources, the gas may be hotter
near to more luminous stars.  Higher mass stars have
a larger temperature difference between hot gas and warm dust, and
also a larger spatial separation of the two components, producing
larger temperature gradients.  These larger gradients, in turn, lead
to larger slopes in the $V^2$ and fluxes versus wavelength.

This is not to say that T Tauri stars do not have significant gaseous
emission.  Models including gaseous emission provide 
superior fits to our data, and previous authors have argued that gaseous
emission is needed to self-consistently model SEDs and visibilities
\citep{AKESON+05}.  Furthermore, modeling of observed gaseous emission line 
profiles from T Tauri stars under the assumption of Keplerian rotation has 
provided evidence for gaseous emission interior to the dust sublimation 
radius \citep[e.g.,][]{NAJITA+06}.  
However, the observed temperature gradients between
dust and gas appear less pronounced than do those around higher-mass stars.

\subsection{The Nature of the Gaseous Opacity \label{sec:opacity}}
Water is predicted to be abundant in the inner regions of disks 
\citep[e.g.,][]{GH04}, although it can be reduced in abundance in highly 
irradiated disks \citep[e.g.,][]{TB05}.  \citet{MUZEROLLE+04} predict that 
dust-free inner disks will commonly show water in emission or absorption 
depending on the disk accretion rate \citep[see also][]{CHK91}.
However, near-IR water spectral features are detected only rarely in T Tauri 
stars \citep[e.g.,][]{CTN04,NAJITA+06} and have only been reported in a few 
Herbig Ae/Be stars \citep[][Najita et al. 2008]{TB05}.

For most sources, a model that includes water vapor can 
fit the data well.  For AS 205 A, it appears that water vapor alone can 
reproduce  the observed gaseous emission interior to the dust sublimation 
radius.  This is consistent with the detection of abundant water vapor, and an 
inferred origin at stellocentric radii as small as 0.3 AU, in high dispersion 
$L$-band spectra of AS 205 \citep{SALYK+08}.

However, in some cases (e.g., for RY Tau), models including
water vapor provide fits of lesser quality than models where the gas
emits only continuum radiation.  Furthermore, for all sources except AS 205 A,
our best-fit models require continuum emission in addition to water vapor
emission, since the gas is inferred to be hotter than 3000 K in the 
innermost regions.  This casts some doubt as to whether H$_2$O emission
is needed to fit the data in all cases.

For the columns of water vapor implied by our best-fit models, one might
expect our targets to show H$_2$O emission lines in lower spatial resolution
spectroscopic data.  High dispersion spectroscopic observations of several
of our sample objects in 
the $K$-band failed to detect strong water emission lines 
\citep[Najita et al. 2008;][]{MANDELL+08},
and archival ISO/SWS data for many of our targets do not show 
longer-wavelength H$_2$O features that typically accompany the 2 $\mu$m
features modeled in this work. These null results provide
further reason to question whether water vapor is a viable 
explanation for the observed compact, hot, circumstellar continuum emission 
from some of our sample.

We therefore pose the question: what can produce gaseous
continuum emission in the inner regions of protoplanetary disks?
We discuss several possible explanations below, and argue that emission 
from free-free transitions of Hydrogen and from the negative Hydrogen ion 
is the most viable mechanism.

\subsubsection{Refractory Dust Grains \label{sec:dustopac}}
While dust grains produce continuum emission, the inferred temperatures
of the hot component ($>3000$ K) in our data are higher than the sublimation
temperatures for even the most refractory dust grains 
\citep[e.g.,][]{POLLACK+94}.  Even for calcium, magnesium, or titanium rich 
oxides, sublimation temperatures above 2000 K occur only for ambient pressures
higher than 0.1--1 
bar \citep[e.g.,][]{LEWIS97}, orders of magnitude larger than 
the pressures expected in protoplanetary disks 
\citep[$<0.01$ bar for typical densities and temperatures in 
inner disk regions; e.g.,][]{MUZEROLLE+04}. 
Dust emission therefore seems highly implausible
as an explanation for the hot, compact emission.

\subsubsection{High$-n$ Atomic Hydrogen \label{sec:highn}}
Recombination of photoionized hydrogen into high-$n$ states may also produce a 
(pseudo) continuum opacity.   This mechanism is also probably responsible
for the Br$\gamma$ emission observed toward many of our targets, since it
produces hydrogen in the $n=7$ level, which can then cascade down to the
$n=4$ level and produce the line emission.  Even higher$-n$ states, which
are longer-lived, might be able to produce more extended and
continuum-like emission.  Photoionization cross sections for these 
high-$n$ states are $\sim 10^{-18}$ cm$^{-2}$, and so assuming
a gas column density of $\sim 10^{27}$ cm$^{-2}$ \citep[a typical value for
young stars accreting material at $\sim 10^{-8}$ M$_{\odot}$ yr$^{-1}$; e.g.,]
[]{MUZEROLLE+04},
we see that a fractional abundance of these high$-n$ states of 
$\ga 10^{-9}$ would be needed to produce an optical depth larger than unity.
Estimating the fractional abundance requires knowledge of the ionization and
radiation structure of the disk, which is beyond the scope of this paper.
However, it seems plausible that recombination to high$-n$ hydrogen states
could contribute some opacity in the inner disk.

\subsubsection{Free-free \label{sec:ff}}
Free-free radiation, produced in ionized winds or accretion columns, or in
disks (as for classical Be stars), may also produce continuum emission.  
Free-free emission may occur around
the higher-mass stars in our sample, which produce more
intense ionizing radiation fields, or in shocked, infalling gas
around lower-mass stars.  The cross section of free-free emission at
wavelengths around 2 $\mu$m is $\sim 10^{-25}$ cm$^4$ dyne$^{-1}$ at 
temperatures of a few thousand Kelvin.  Assuming the free-free emission arises
in a disk with a gas temperatuture of 3000 K, a gas
column density of $10^{27}$ cm$^{-2}$, a fractional ionization of $10^{-5}$,
and H$/$H$_2$=0.01, we find an electron pressure
of $\sim 0.01$ dyne cm$^{-2}$, and an optical depth of $\sim 0.01$.
Fractional ionizations of $10^{-5}$ can arise from the ionization of metals
with low ionization potentials, such as Na or Fe; higher fractional ionizations
are difficult to achieve since photons with $>11$ eV are unlikely to penetrate
deeply into  such a dense disk.  However, somewhat higher optical depths of 
free-free emission may be achieved if the gas column density is higher than the
value assumed above.  Some of our targets may be accreting faster than
$10^{-8}$ M$_{\odot}$ yr$^{-1}$, leading to higher gas columns and 
more free-free emission.  It thus seems possible that free-free emission
could contribute significantly in the inner disk.

\subsubsection{H$^{-}$ \label{sec:hminus}}
The negative hydrogen ion, H$^-$, appears as a promising candidate for
explaining this emission, since it can produce free-free continuum emission
in the $K$ band  \citep[e.g.,][]{CB46,BB87,JOHN88,GRAY92}. 
For free-free 
emission from H$^-$ at $\sim 3000$ K in the $K$-band, the optical depth is 
\citep[following][]{GRAY92} 
\begin{equation} 
\tau \sim 10^{-38} \left(\frac{n_e}{\rm cm^{-3}}\right)
\left(\frac{N_H}{\rm cm^{-2}}\right).
\end{equation}
With the same assumptions at in \S \ref{sec:ff}, we find an optical depth
of $\sim 0.01$ in the inner disk.  As above, higher gas columns may lead
to correspondingly higher optical depths.   Free-free
emission from H$^-$ therefore seems like a plausible explanation 
for the hot, compact emission seen in our data.

\section{Conclusions \label{sec:conc}}
We presented spatially resolved near-IR spectroscopic observations that 
probed the gas and dust in the inner disks around 15 young stars.
One source, HD 141569, was unresolved at all wavelengths between 2.0 and 2.4 
$\mu$m, indicating a lack of dust or gas in the inner disk regions.  Another
target, VV Ser, was over-resolved, indicating that the emission spans angles
larger than $\sim 5$ mas at all observed wavelengths.  

The near-IR emission
from the remaining targets was resolved, and our data show that the angular 
size of the near-IR emission increases with wavelength in all cases.
This behavior suggests temperature gradients in these inner disks,
arising from the combination of warm dust at its sublimation
temperature and hotter, presumably gaseous material within the dust sublimation
radius.  Our data clearly indicate emission from the Br$\gamma$ transition of 
hydrogen in several objects, and suggest that 
water vapor and carbon monoxide gas are present 
in the inner disks of some targets.  

We constructed simple physical models of the inner disk, including dust
and gas emission, and we fitted them to our data to constrain the spatial 
distribution and temperature of dust and gas emission components.  We 
considered models including only dust emission; dust, gas continuum, and
Br$\gamma$ emission; and dust, gas continuum, water vapor, and Br$\gamma$
emission.  Models incorporating only dust emission can not fit the data 
for any of our sources well.  In contrast, 
models including dust and gas emission
are suitable for explaining our data.  
The inclusion or exclusion of water vapor
in these dust+gas models did not substantially affect the quality of
the fits in most cases.

For all sources where Br$\gamma$ emission is observed, we find it to be compact
relative to the continuum emission.  This contrasts with previous 
findings, which found Br$\gamma$ emission to be extended relative to
the continuum around some high-mass stars.  The results presented here
suggest that Br$\gamma$ commonly traces infalling material around young stars 
spanning a large range in stellar mass. 

CO emission is tentatively 
observed towards several objects, and we see evidence that this emission
has a more compact spatial distribution than the dust around RW Aur.  For
other objects, our data are insufficient to place meaningful constraints
on the relative spatial distribution of CO and other emission components.
We will re-observe these targets in the near future with higher dispersion,
to obtain better signal-to-noise for the relatively narrow CO lines
and better constrain their spatial distribution.

While models including water vapor opacity often fit our data well, 
the best-fit models generally also require continuum emission from material
that is too hot to be water (since water dissociates at $\sim 3000$ K).
We do not have a ready explanation for the source of this hot continuum
emission, but we speculate that it may trace free-free emission from hydroden
and/or H$^-$.  The gas densities and fractional ionizations required to 
produce such emission seem plausible in the inner regions of protoplanetary 
disks, suggesting that free-free emission from H and H$^-$ is a viable 
explanation for the compact continuum emission seen in our data.

\medskip

Data presented herein were obtained at the W. M. Keck Observatory, in part 
from telescope time allocated to the National Aeronautics and Space 
Administration through the agency's scientific partnership with the California 
Institute of Technology and the University of California. The Observatory was 
made possible by the generous financial support of the W. M. Keck Foundation. 
The authors wish to recognize and acknowledge the cultural role and reverence
that the summit of Mauna Kea has always had within the indigenous Hawaiian
community. We are most fortunate to have the opportunity to conduct 
observations from this mountain. This work has used software from the 
Michelson Science Center at the California Institute of Technology.
The authors thank the entire Keck Interferometer team for making 
these observations possible.  We also wish to thank the referee, Geoff Blake,
for his thoughtful and detailed referee report, which greatly improved the
manuscript.

\clearpage
\begin{deluxetable}{lccccccc}
\tabletypesize{\scriptsize}
\tablewidth{0pt}
\tablecaption{Target and Calibrator Properties
\label{tab:sample}}
\tablehead{\colhead{Source} & \colhead{$\alpha$} 
& \colhead{$\delta$} & \colhead{$d$} & \colhead{Spectral Type} & 
\colhead{$m_{V}$} & \colhead{$m_{K}$} & \colhead{References}}
\startdata
\multicolumn{8}{c}{Target Stars} \\
\hline
RY Tau & 04 21 57.409 & +28 26 35.56 & 140 & K1 & 10.2 & 5.4 & 1 \\
DG Tau & 04 27 04.700 & +26 06 16.20 & 140 & K3 & 12.4 & 7.0 & 2 \\
MWC 480 & 04 58 46.266 & +29 50 37.00 & 140 & A2 & 7.7 & 5.5 & 3 \\ 
RW Aur A & 05 07 49.568 & +30 24 05.161 & 140 & K2 & 10.5 & 7.0 & 2 \\  
MWC 758 & 05 30 27.530 & +25 19 57.08 & 140 & A3 & 8.3 & 5.8 & 3 \\
HD 141569 & 15 49 57.75 & -03 55 16.4 & 99 & B9/A0 & 7.1 & 6.7 & 3 \\
HD 142666 & 15 56 40.023 & -22 01 40.01 & 116 & A8 & 8.8 & 6.1 & 4 \\
HD 144432 & 16 06 57.957  & -27 43 09.81 & 145 & A9 & 8.2 & 5.9 & 4 \\
AS 205 A & 16 11 31.402 & -18 38 24.54 & 160 & K5 & 12.1 & 6.0 & 5 \\
MWC 863 A & 16 40 17.922 & -23 53 45.18 & 150 & A2 & 8.9 & 5.5 & 4 \\
MWC 275 & 17 56 21.288 & -21 57 21.88 & 122 & A1 & 6.9 & 4.8 & 4 \\
VV Ser & 18 28 47.860 & +00 08 40.00 & 310 & A0 & 11.9 & 6.3 & 3 \\
V1295 Aql & 20 03 02.510 & +05 44 16.68 & 290 & B9/A0 & 7.8 & 5.9 & 3 \\
AS 442 & 20 47 37.470 & +43 47 24.90 & 600 & B8 & 10.9 & 6.6 & 3 \\
MWC 1080 & 23 17 25.574 & +60 50 43.34 & 1000 & B0 & 11.6 & 4.7 & 3 \\
\hline
\multicolumn{7}{c}{Calibrator Stars} & Applied to: \\
\hline
HD 27777 & 04 24 29.155 & +34 07 50.73 & 187 & B8V & 5.7 & 6.0 & RY Tau,MWC 480,MWC 758 \\
HD23642 & 03 47 29.453 & +24 17 18.04 &  110 & A0V & 6.8 & 6.8 & DG Tau,RW Aur \\
HD23632 & 03 47 20.969 & +23 48 12.05 & 120 & A1V & 7.0 & 7.0 & DG Tau,RW Aur \\
%HD283668 & 04 27 52.933 & +24 26 41.24 & 42 & K3V & 9.4 & 7.0 & DG Tau \\
HD31464 & 04 57 06.426 & +24 45 07.90 & 45 & G5V & 8.6 & 7.0 & DG Tau,RW Aur \\
HD139364 & 15 38 25.358 & -19 54 47.45 & 53 & F3V & 6.7 & 5.7 & HD 142666,HD 144432,AS 205 A \\
HD 141247 & 15 48 11.770 & -04 47 09.684 & 85 & F9V & 8.1 & 6.7 & HD  141569 \\
HD 143459 & 16 00 47.633 & -08 24 40.87 & 141 & A0Vs & 5.5 & 5.5 & HD 142666,HD 144432,AS 205 A \\
HD 145788 & 16 12 56.583 & -04 13 14.912 & 171 & A1V & 6.3 & 6.2 & HD 141569 \\
HD 149013 & 16 32 38.133 & -15 59 15.12 & 41 & F8V & 7.0 & 5.7 & MWC 863 \\
HD 163955 & 17 59 47.553 & -23 48 58.08 & 134 & B9V & 4.7 & 4.9 & MWC 275 \\
HD 170657 & 18 31 18.960 & -18 54 31.72 & 13 & K1V & 6.8 & 4.7 & MWC 275 \\
HD 183324 & 19 29 00.988 & +01 57 01.61 & 59 & A0V & 5.8 & 5.8 & V1295 Aql \\
HD 188385 & 19 54 40.200 & +07 08 25.27 & 81 & A2V & 6.1 & 6.1 & V1295 Aql \\
HD199099 & 20 53 26.390 & +42 24 36.72 & 138 & A1V & 6.7 & 6.7 & AS 442,MWC 1080 \\
%HD219290 & 23 14 14.385 & +50 37 04.41 & 00 & A0V & 6.3 & 6.3 & AS 442,MWC 1080 \\
HD1404 & 00 18 19.657 & +36 47 06.81 & 43 & A2V & 4.5 & 4.5 & AS 442,MWC 1080 \\
\enddata
\tablerefs{(1) \citet{MUZEROLLE+03}; (2) \citet{WG01}; (3) \citet{EISNER+04};
(4) \citet{MONNIER+06}; (5) \citet{EISNER+05}.  Calibrator star distances are
based on Hipparcos parallax measurements \citep{PERRYMAN+97}.}
\end{deluxetable}

\begin{deluxetable}{lcccccc}
\tabletypesize{\scriptsize}
\tablewidth{0pt}
\tablecaption{Results of Modeling
\label{tab:results}}
\tablehead{\colhead{Source}
& \colhead{$\chi_r^2$} & \colhead{$T_{\rm in}$} & \colhead{$R_{\rm in}$} & 
\colhead{$\tau_{\rm in}$} & \colhead{$T_{\rm dust}$} & \colhead{$R_{\rm dust}$} \\
 & & (K) & (AU) & & (K) & (AU)}
\startdata
\multicolumn{7}{c}{Dust Models} \\
\hline
RY Tau  &  44.34  &  &  &  &  1750 $\pm$ 10  & 0.16 $\pm$ 0.01 \\
MWC 480  &  31.85  &  &  &  &  1450 $\pm$ 10  & 0.20 $\pm$ 0.01 \\
MWC 758  &  17.89  &  &  &  &  1610 $\pm$ 10  & 0.17 $\pm$ 0.01 \\
DG Tau  &  4.70  &  &  &  &  1260 $\pm$ 10  & 0.18 $\pm$ 0.01 \\
RW Aur  &  3.60  &  &  &  &  1330 $\pm$ 10  & 0.14 $\pm$ 0.01 \\
HD 142666  &  1.55  &  &  &  &  1670 $\pm$ 10  & 0.10 $\pm$ 0.01 \\
HD 144432  &  10.92  &  &  &  &  1420 $\pm$ 10  & 0.20 $\pm$ 0.01 \\
MWC 863A  &  41.40  &  &  &  &  1360 $\pm$ 10  & 0.28 $\pm$ 0.01 \\
V1295 Aql  &  25.20  &  &  &  &  1350 $\pm$ 10  & 0.48 $\pm$ 0.01 \\
AS 205A  &  2.08  &  &  &  &  1850 $\pm$ 10  & 0.14 $\pm$ 0.01 \\
MWC 275  &  19.93  &  &  &  &  1750 $\pm$ 10  & 0.20 $\pm$ 0.01 \\
AS 442  &  8.63  &  &  &  &  1580 $\pm$ 10  & 0.57 $\pm$ 0.01 \\
MWC 1080  &  24.39  &  &  &  &  2190 $\pm$ 10  & 1.39 $\pm$ 0.01 \\
\hline
\multicolumn{7}{c}{Dust+Gas Continuum Models} \\
\hline
RY Tau  &  0.30  &  2684 $\pm$ 371  & 0.02 $\pm$ 0.04 &  1.623 $\pm$ 4.655  &  1105 $\pm$ 110  & 0.31 $\pm$ 0.02 \\
MWC 480  &  1.04  &  3199 $\pm$ 391  & 0.07 $\pm$ 0.01 &  0.081 $\pm$ 0.029  &  1105 $\pm$ 9  & 0.28 $\pm$ 0.01 \\
MWC 758  &  1.73  &  3462 $\pm$ 167  & 0.07 $\pm$ 0.01 &  0.104 $\pm$ 0.013  &  849 $\pm$ 53  & 0.35 $\pm$ 0.03 \\
DG Tau  &  1.48  &  2188 $\pm$ 423  & 0.01 $\pm$ 0.01 &  2.000 $\pm$ 0.001  &  1137 $\pm$ 29  & 0.21 $\pm$ 0.01 \\
RW Aur  &  1.71  &  7125 $\pm$ 981  & 0.02 $\pm$ 0.01 &  0.009 $\pm$ 0.001  &  1233 $\pm$ 26  & 0.15 $\pm$ 0.01 \\
HD 142666  &  0.13  &  5577 $\pm$ 22197  & 0.01 $\pm$ 0.12 &  0.175 $\pm$ 0.989  &  1236 $\pm$ 736  & 0.14 $\pm$ 0.07 \\
HD 144432  &  0.54  &  3910 $\pm$ 479  & 0.07 $\pm$ 0.01 &  0.048 $\pm$ 0.015  &  1031 $\pm$ 19  & 0.29 $\pm$ 0.01 \\
MWC 863A  &  0.67  &  7714 $\pm$ 88  & 0.07 $\pm$ 0.01 &  0.009 $\pm$ 0.001  &  1128 $\pm$ 5  & 0.35 $\pm$ 0.01 \\
V1295 Aql  &  0.58  &  3115 $\pm$ 464  & 0.14 $\pm$ 0.01 &  0.063 $\pm$ 0.029  &  1106 $\pm$ 10  & 0.62 $\pm$ 0.01 \\
AS 205A  &  0.58  &  6952 $\pm$ 3684  & 0.01 $\pm$ 0.01 &  2.000 $\pm$ 0.001  &  1461 $\pm$ 68  & 0.18 $\pm$ 0.01 \\
MWC 275  &  0.13  &  3983 $\pm$ 544  & 0.06 $\pm$ 0.01 &  0.129 $\pm$ 0.045  &  1230 $\pm$ 14  & 0.28 $\pm$ 0.01 \\
AS 442  &  2.70  &  2521 $\pm$ 924  & 0.29 $\pm$ 0.04 &  0.199 $\pm$ 0.282  &  681 $\pm$ 1688  & 1.34 $\pm$ 4.36 \\
MWC 1080  &  0.67  &  2790 $\pm$ 218  & 0.50 $\pm$ 0.01 &  0.687 $\pm$ 0.173  &  1271 $\pm$ 14  & 2.33 $\pm$ 0.04 \\
\hline
\multicolumn{7}{c}{Dust+Water Models} \\
\hline
RY Tau  &  2.21  &  5427 $\pm$ 654  & 0.02 $\pm$ 0.01 &  0.238 $\pm$ 0.055  &  1211 $\pm$ 4  & 0.26 $\pm$ 0.09 \\
MWC 480  &  2.46  &  4659 $\pm$ 799  & 0.02 $\pm$ 0.01 &  0.173 $\pm$ 0.084  &  1246 $\pm$ 3  & 0.25 $\pm$ 0.10 \\
MWC 758  &  2.41  &  4632 $\pm$ 738  & 0.02 $\pm$ 0.01 &  0.201 $\pm$ 0.064  &  1241 $\pm$ 4  & 0.24 $\pm$ 0.08 \\
DG Tau  &  1.56  &  16907 $\pm$ 470  & 0.01 $\pm$ 0.06 &  0.020 $\pm$ 0.273  &  1204 $\pm$ 18  & 0.19 $\pm$ 0.09 \\
RW Aur  &  1.68  &  6114 $\pm$ 2414  & 0.03 $\pm$ 0.01 &  0.009 $\pm$ 0.000  &  1226 $\pm$ 25  & 0.15 $\pm$ 0.01 \\
HD 142666  &  0.25  &  7334 $\pm$ 6398  & 0.01 $\pm$ 0.01 &  0.757 $\pm$ 1.496  &  1396 $\pm$ 5  & 0.12 $\pm$ 0.28 \\
HD 144432  &  1.13  &  6812 $\pm$ 3310  & 0.01 $\pm$ 0.01 &  0.574 $\pm$ 0.711  &  1225 $\pm$ 2  & 0.25 $\pm$ 0.24 \\
MWC 863A  &  2.55  &  5090 $\pm$ 1149  & 0.01 $\pm$ 0.01 &  0.187 $\pm$ 0.103  &  1218 $\pm$ 2  & 0.33 $\pm$ 0.14 \\
V1295 Aql  &  1.68  &  5368 $\pm$ 1560  & 0.02 $\pm$ 0.01 &  0.232 $\pm$ 0.053  &  1222 $\pm$ 1  & 0.55 $\pm$ 0.32 \\
AS 205A  &  0.48  &  4222 $\pm$ 2727  & 0.01 $\pm$ 0.01 &  0.611 $\pm$ 0.602  &  1631 $\pm$ 6  & 0.16 $\pm$ 0.14 \\
MWC 275  &  0.57  &  5271 $\pm$ 897  & 0.02 $\pm$ 0.01 &  0.246 $\pm$ 0.059  &  1409 $\pm$ 4  & 0.25 $\pm$ 0.19 \\
AS 442  &  2.92  &  13212 $\pm$ 14239  & 0.01 $\pm$ 0.01 &  2.093 $\pm$ 7.427  &  1375 $\pm$ 3  & 0.67 $\pm$ 2.54 \\
MWC 1080  &  1.39  &  10730 $\pm$ 2952  & 0.06 $\pm$ 0.04 &  0.782 $\pm$ 0.750  &  1600 $\pm$ 9  & 1.90 $\pm$ 3.33 \\
\enddata
\end{deluxetable}

\begin{deluxetable}{l|cc|cc}
\tabletypesize{\scriptsize}
\tablewidth{0pt}
\tablecaption{Properties of Br$\gamma$ Emission
\label{tab:brg}}
\tablehead{\colhead{$ $} & \multicolumn{2}{c}{Dust+Gas Model} & 
\multicolumn{2}{c}{Dust+Water Model} \\
\colhead{Source} & \colhead{$R_{\rm Br \gamma}$} & 
\colhead{$F_{\rm Br\gamma}/F_{\rm cont}$} & \colhead{$R_{\rm Br \gamma}$} & 
\colhead{$F_{\rm Br\gamma}/F_{\rm cont}$} \\
 & (AU) &  & (AU) & }
\startdata
RY Tau  & 0.02 $\pm$ 0.04 &  0.01 $\pm$ 0.03 & 0.07 $\pm$ 0.01 &  0.02 $\pm$ 0.02 \\
MWC 480  & 0.07 $\pm$ 0.01 &  0.08 $\pm$ 0.02 & 0.01 $\pm$ 0.01 &  0.09 $\pm$ 0.03 \\
MWC 758  & 0.02 $\pm$ 0.01 &  0.03 $\pm$ 0.03 & 0.01 $\pm$ 0.16 &  0.05 $\pm$ 0.02 \\
DG Tau  & 0.07 $\pm$ 0.01 &  0.16 $\pm$ 0.04 & 0.07 $\pm$ 0.01 &  0.16 $\pm$ 1.13 \\
RW Aur  & 0.07 $\pm$ 0.01 &  0.06 $\pm$ 0.04 & 0.01 $\pm$ 0.01 &  0.06 $\pm$ 0.05 \\
HD 142666  & 0.05 $\pm$ 0.01 &  0.01 $\pm$ 0.05 & 0.05 $\pm$ 0.01 &  0.02 $\pm$ 0.05 \\
HD 144432  & 0.07 $\pm$ 0.01 &  0.04 $\pm$ 0.03 & 0.02 $\pm$ 0.01 &  0.05 $\pm$ 0.05 \\
MWC 863A  & 0.02 $\pm$ 0.01 &  0.04 $\pm$ 0.01 & 0.02 $\pm$ 0.01 &  0.05 $\pm$ 0.02 \\
V1295 Aql  & 0.01 $\pm$ 0.12 &  0.08 $\pm$ 0.02 & 0.06 $\pm$ 0.02 &  0.09 $\pm$ 0.02 \\
AS 205A  & 0.01 $\pm$ 0.01 &  0.01 $\pm$ 0.01 & 0.01 $\pm$ 0.01 &  0.02 $\pm$ 0.04 \\
MWC 275  & 0.01 $\pm$ 0.01 &  0.05 $\pm$ 0.02 & 0.01 $\pm$ 0.01 &  0.06 $\pm$ 0.02 \\
AS 442  & 0.05 $\pm$ 0.05 &  0.04 $\pm$ 0.19 & 0.12 $\pm$ 0.03 &  0.03 $\pm$ 0.06 \\
MWC 1080  & 0.01 $\pm$ 0.02 &  0.05 $\pm$ 0.02 & 0.06 $\pm$ 0.04 &  0.04 $\pm$ 0.02 \\
\enddata
\tablecomments{The uncertainties listed here are 1$\sigma$ statistical errors
for our model fits.  As discussed in \S \ref{sec:resmain}, these error bars are
probably too small for $R_{\rm Br \gamma}$, although we can state confidently
that $R_{\rm Br \gamma} < 0.1$ AU for these sources.}
\end{deluxetable}

\clearpage

\epsscale{0.8}

\epsscale{1.0}

%\epsscale{1.0}
%\begin{figure}
%\plotone{figs/bblaw.eps}
%\caption{Flux versus radius (and temperature) for a simple disk model.  
%We plot the predictions for an optically thick disk ({\it solid curve})
%and disks where we assume that the optical depth 
%decreases radially as $r^{-3/2}$ ({\it dotted and dashed curves}).  
%This model assumes that the central star is a 2 M$_{\odot}$ star with 
%a radius of 2 R$_{\odot}$, an effective temperature of 9000 K, and an
%accretion rate of $10^{-7}$ M$_{\odot}$ yr$^{-1}$.  The central source
%then heats a geometrically thin disk, yielding a temperature profile
%$T \propto R^{-1/2}$. We allow this disk to extend from 0.025 AU, expected
%to be within the magnetospheric truncation radius, out to 0.25 AU,
%where disk temperatures become suitable for the existence of dust.
%\label{fig:bblaw}}
%\end{figure}

\begin{figure}
\plotone{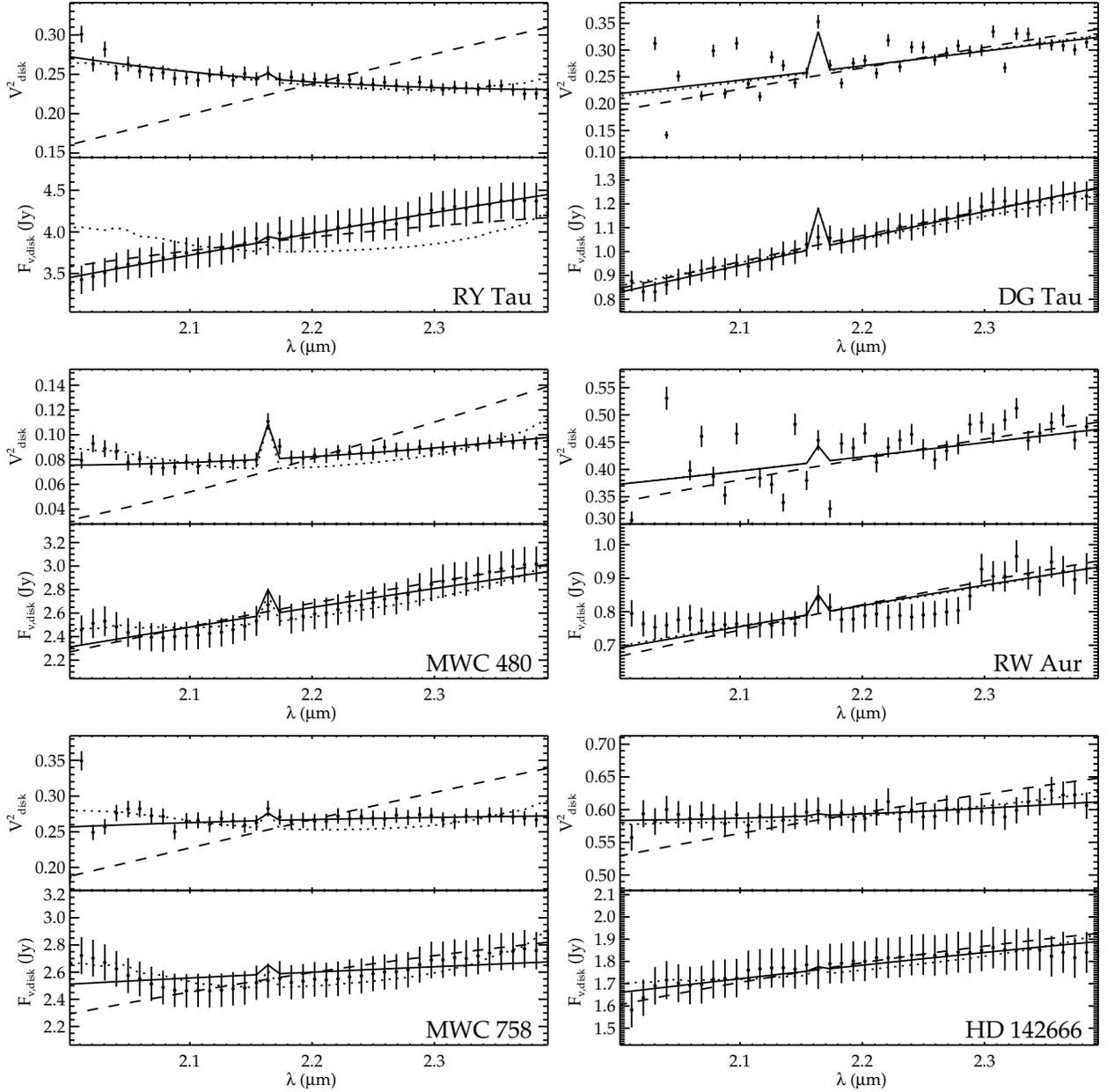}
\caption{The circumstellar 
components of measured $V^2$ and fluxes for our sample,
as a function of wavelength, with the predictions of various models.
Dashed curves show the predictions of models including a single-temperature
ring of (dust) emission.  Solid curves represent a model that includes 
a gaseous inner disk that emits continuum emission and Br$\gamma$ emission,
in addition to the dust ring.  Dotted curves 
show the predictions of models that include the effects of water vapor
opacity in the gaseous emission.  
\label{fig:fits}}
\end{figure}

\clearpage
\epsscale{1.0}
{\plotone{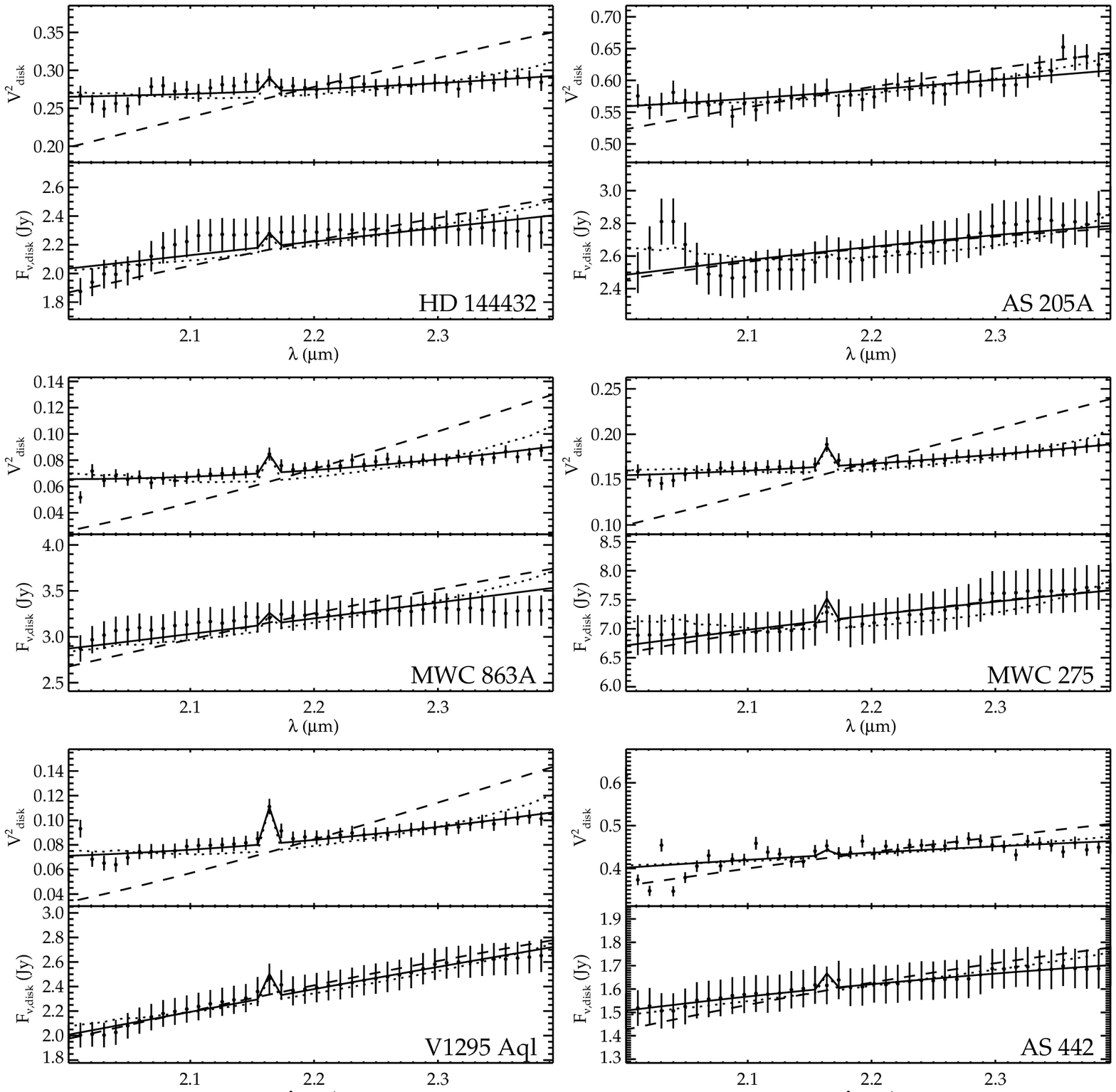}}\\[5mm]
\centerline{Fig. \ref{fig:fits}. --- continued.}
\clearpage

\clearpage
\epsscale{1.0}
{\plotone{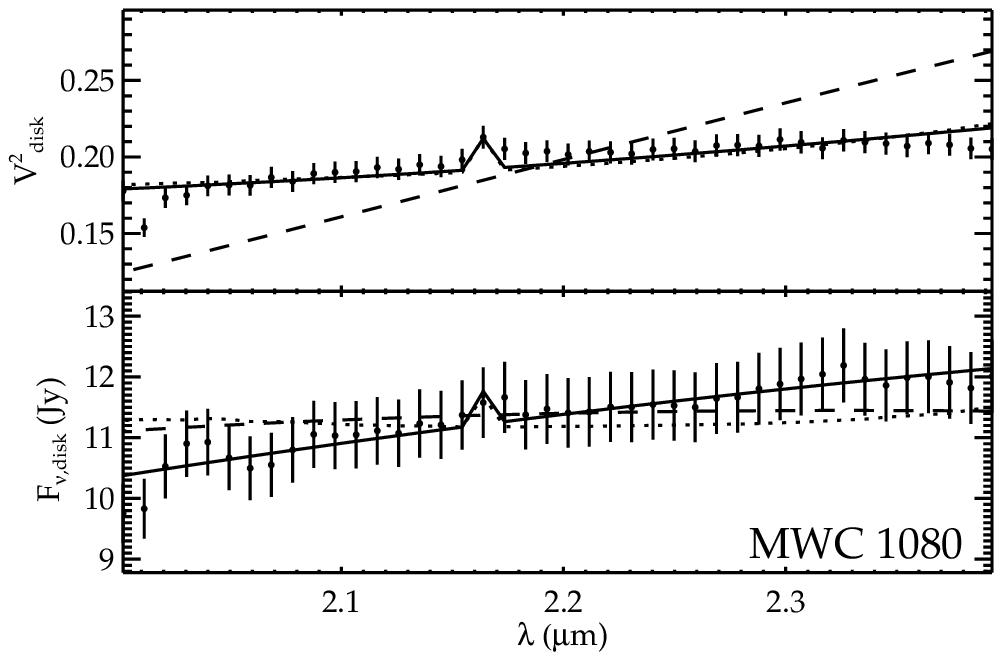}}\\[5mm]
\centerline{Fig. \ref{fig:fits}. --- continued.}
\clearpage

\epsscale{0.8}

\end{document}